\documentclass[%
longbibliography,
amsmath,amssymb,
aps,
pra,
]{revtex4-1}
\usepackage{graphicx}
\usepackage{dcolumn}
\usepackage{bm}
\usepackage{epsfig}
\usepackage{afterpage}

\begin{document}

\title{Artificial Magnetic Field Quenches in Synthetic Dimensions}

\author{F. Y{\i}lmaz}
\email{firat.yilmaz@bilkent.edu.tr}
\author{M. \"{O}. Oktel}
\affiliation{Department of Physics, Bilkent University, Ankara 06800, Turkey}
\date{\today}

\begin{abstract}

Recent cold atom experiments have realized models where each hyperfine state at an optical lattice site can be regarded as a separate site in a synthetic dimension. In such synthetic ribbon configurations, manipulation of the transitions between the hyperfine levels provide direct control of the hopping in the synthetic dimension. This effect  was used to simulate a magnetic field through the ribbon. Precise control over the hopping matrix elements in the synthetic dimension makes it possible to change this artificial magnetic field much faster than the time scales associated with atomic motion in the lattice. In this paper, we consider such a magnetic flux quench scenario in synthetic dimensions. Sudden changes have not been considered for real magnetic fields as such changes in a conducting system would result in large induced currents. Hence, we first study the difference between a time varying real magnetic field and an artificial magnetic field using a minimal six site model. This minimal model clearly shows the connection between gauge dependence and the lack of on site induced scalar potential terms. We then investigate the dynamics of a wavepacket in an infinite two or three leg ladder following a flux quench and find that the gauge choice has a dramatic effect on the packet dynamics. Specifically, a wavepacket splits into a number of smaller packets moving with different velocities. Both the weights and the number of packets depend on the implemented gauge. If an initial packet, prepared under zero flux in a n--leg ladder, is quenched to Hamiltonian with a vector potential parallel to the ladder; it splits into at most $n$ smaller wavepackets. The same initial wavepacket splits into up to $n^2$ packets if the vector potential is implemented to be along the rungs. Even a trivial difference in the gauge choice such as the addition of a constant to the vector potential produces observable effects. We also calculate the packet weights for arbitrary initial and final fluxes. Finally, we show that edge states in a thick ribbon are robust under the quench only when the same gap supports an edge state for the final Hamiltonian.
\end{abstract}
\maketitle

\section{Introduction}
Cold atom experiments provide a clean and controlled environment for investigating many body systems. The ability to change the Hamiltonian of a system at much shorter timescales compared to dynamics of the atoms in the experiment, combined with the long lifetime of the excitations in clean systems, have allowed the exploration of non-equilibrium physics in interacting many-body models. Both sudden and periodic changes in system parameters, i.e. quench \cite{BlochQuench,spielmanQuench} and Floquet \cite{FloquetDynTunn,FloquetLignier,FloquetZener} experiments, have revealed non-trivial phenomena such as many body localization\cite{HuseLocalization,ManBodyLocaliBloch}, pre-thermalized states\cite{eisertPreThermalized} and new topological invariants\cite{rudnerTimDepTopInvt}.

In a parallel development, creation of artificial gauge fields in continuum\cite{IanSpielman} and in the lattice \cite{BlochHofstadter,KetterleHarper} paved the way for the study of effectively charged particles with neutral atoms. The physics of particles interacting with abelian and non-abelian artificial gauge fields is rich. Recent experiments have probed the topological properties \cite{BlochChern,GoldmanTopoEdge,EsslingerHaldane,goldman2016topological}, quantum Hall states\cite{SpielmanSuperfluidHall}, Hofstadter's fractal energy spectrum\cite{KetterleHarper,BlochHofstadter} and supersolid-like states in non-abelian fields\cite{superSolids,superSolids2}. Generally artificial magnetic fields are created by utilizing a combination of light-atom interaction to provide a momentum space shift in a spatially varying parameter to create a spatially dependent vector potentials\cite{DalibardReview}. The combination of a momentum-space technique with real space manipulation, especially in lattices, makes these experiments highly non-trivial, and puts technical limitations on the homogeneity and the value of the artificial magnetic field, as well as the lifetime of the system\cite{BlochChern,KetterleHarper}.

Our main interest in this paper is to investigate the effects of a sudden change in the artificial magnetic field in a cold atom system. Until recently, it would not be realistic to expect the artificial magnetic field in a lattice system to be controlled at fast time scales. However, a new method of simulating lattice systems which is based on the identification of hyperfine levels of each atom with a lattice in a 'synthetic dimension' enables, perhaps the simplest method for creation of artificial magnetic fields \citep{GoldmanArtGaugeFieldInSyntheDim}. The hyperfine levels are coupled by Raman lasers such that the Peierls phase is introduced by the detuning. Since a typical Raman transition frequency ($\sim$300 THz) is much larger than typical frequencies of motion ($\sim$100 Hz)  in the system, a non-adiabatic change in the artificial field can be studied, unlike its real space two-dimensional counterparts discussed above. The dynamics of both Bosons and Fermions have been investigated in a static artificial magnetic field\cite{SyntheticChiralEdge,SpielmanEdgeQHE}, quench experiments as considered in this paper are within the reach of current capabilities of several research groups.

Apart from current experimental relevance, the investigation of such time dependent artificial magnetic fields is interesting for two main reasons. First, artificial magnetic fields, as implemented in the current experiments, rely on mimicking the effect of the vector potential $\vec{A}$ but do not create the effect of the scalar potential $\Phi$. In static fields, two vector potentials $\vec{A}_1$ and $\vec{A}_2$ which have the same curl $\vec{B}=\nabla \times \vec{A}_1=\nabla \times \vec{A}_2$ create entirely equivalent Hamiltonians up to a gauge transformation, hence no observable depends on whether the first or the second choice was implemented experimentally. In a dynamical situation, even if two vector potentials have the same curl at all times $\vec{B}(t)=\nabla \times \vec{A}_1(t)=\nabla \times \vec{A}_2(t)$, the Hamiltonians can not be connected by a gauge transformation in the absence of the scalar potential. This simple 'gauge dependence', has non-trivial consequences as evidenced by the expansion image in a recent experiment which measures the gauge dependent canonical momentum distribution\cite{KetterleGaugeDepdnc}. It is important to notice that in contrast to the gauge-like effect of the micromotion in a Floquet system \cite{micromotion} the gauge dependence in a quench experiment is essentially an infinite frequency effect. Two systems which are connected by a gauge transformation before and after the quench moment can yield different physical results.
The second qualitative question that can be investigated is the robustness of edge states under non-adiabatic quench scenarios. Edge states which arise due to topological reasons are robust with respect to perturbations, however such robustness is not guaranteed by any topological argument after a quench.

To answer the above questions, we analyse the quench dynamics of wavepackets under sudden changes in the artificial magnetic field. We use the tight-binding model for lattice systems, and the effect of the magnetic field is introduced by Peierls substitution\cite{Peierls}. Initially, the role of the gauge choice is considered for a minimal model, a six site lattice forming two squares. An arbitrary initial wavefunction is time evolved for two different vector potential choices. The probability distributions of the wavefunctions are compared as a function of time and are found to depend on the particular gauge choice and how fast the artificial magnetic field is ramped from zero to its final value. We show that a time dependent on site potential can be used to make the two vector field choices equivalent, restoring gauge invariance. These results are presented in section II.

In section III, we consider the magnetic flux quench of wavepackets in  two and three-leg ladders. A wavepacket peaked around an eigenstate of the ladder at zero magnetic field splits into smaller wavepackets which move with different velocities after the quench.
We find that the gauge choice is highly decisive in the number and the weights of the smaller packets. In a gauge choice with vector potential parallel to the ladder, an initial wavepacket in an $n$-leg ladder splits into at most $n$ wavepackets, no matter what the magnetic flux value is. Whereas, a quench to a vector potential parallel to the rungs of the ladder yields up to $n^2$ smaller packets for the same initial wavepacket. The number of wavepackets, as well as their weights and velocities, are found by calculating the overlaps of the wavefunction before the quench with the post-quench bands.
For the two leg ladder, we also consider a quench between two arbitrary magnetic fluxes for the vector potential along the ladder and provide analytical expressions for the packet weights. In addition, we investigate a special case in the three-leg ladder where an initial wavepacket at non-zero flux is quenched to Hamiltonian with zero flux but a constant vector potential. We find that the final value of the vector potential does not change the number or the weights of the split wave packets but modifies the internal hyperfine composition of the packets.

We present our results on the behavior of an edge state after an artificial magnetic field quench in section IV. We investigate both a thick ribbon, which we take to have fifteen legs, and also the continuum problem. For the lattice system the topologically protected edge state survives the quench only if there is another edge state in the same gap after the quench. Otherwise, for example when the final magnetic flux is zero, the edge state disperses into the bulk. In the continuum, the existence of edge modes is guaranteed by the termination of the Landau levels at the edge and the edge modes survive as skipping orbits after a quench. We calculate the trajectories of the center of mass coordinate of a wavepacket to validate that an initial edge state wavepacket is replaced by a skipping trajectory and remains bound to the edge.

Section V contains a summary of our conclusions and their experimental relevance.

\section{A minimal model}
Physics is gauge invariant. How a gauge choice is related to the wavefunction is clearly understood for real scalar and vector potentials. In the case of artificial gauge fields in cold atom experiments, two difficulties complicate the interpretation. First, if the artificial fields change with time, there is no set of equations they have to follow such as Maxwell's equations which dictates their dynamics\cite{ErichMuellerElectro}. As such, a changing artificial magnetic flux does not necessarily amount to an artificial electric field. The second difficulty is that the presence of these fields is only measured through their action on the atoms. For example, the expansion images in the MIT experiment\cite{KetterleGaugeDepdnc} show a density which is proportional to the canonical momentum, which is a manifestly gauge dependent quantity. This measurement can be interpreted in two distinct ways. One can say that the experiment implements the vector potential, but not the scalar potential, hence the simulated artificial fields do not have gauge symmetry. The second, and equally valid, explanation would be to say that a turn-off of the optical lattice is not simply magnetic flux being shut down, but it involves a momentary artificial electric field breaking the $C_4$ rotational symmetry of the lattice.

\begin{figure}
\includegraphics[width=0.7\textwidth,height=0.4\textwidth]{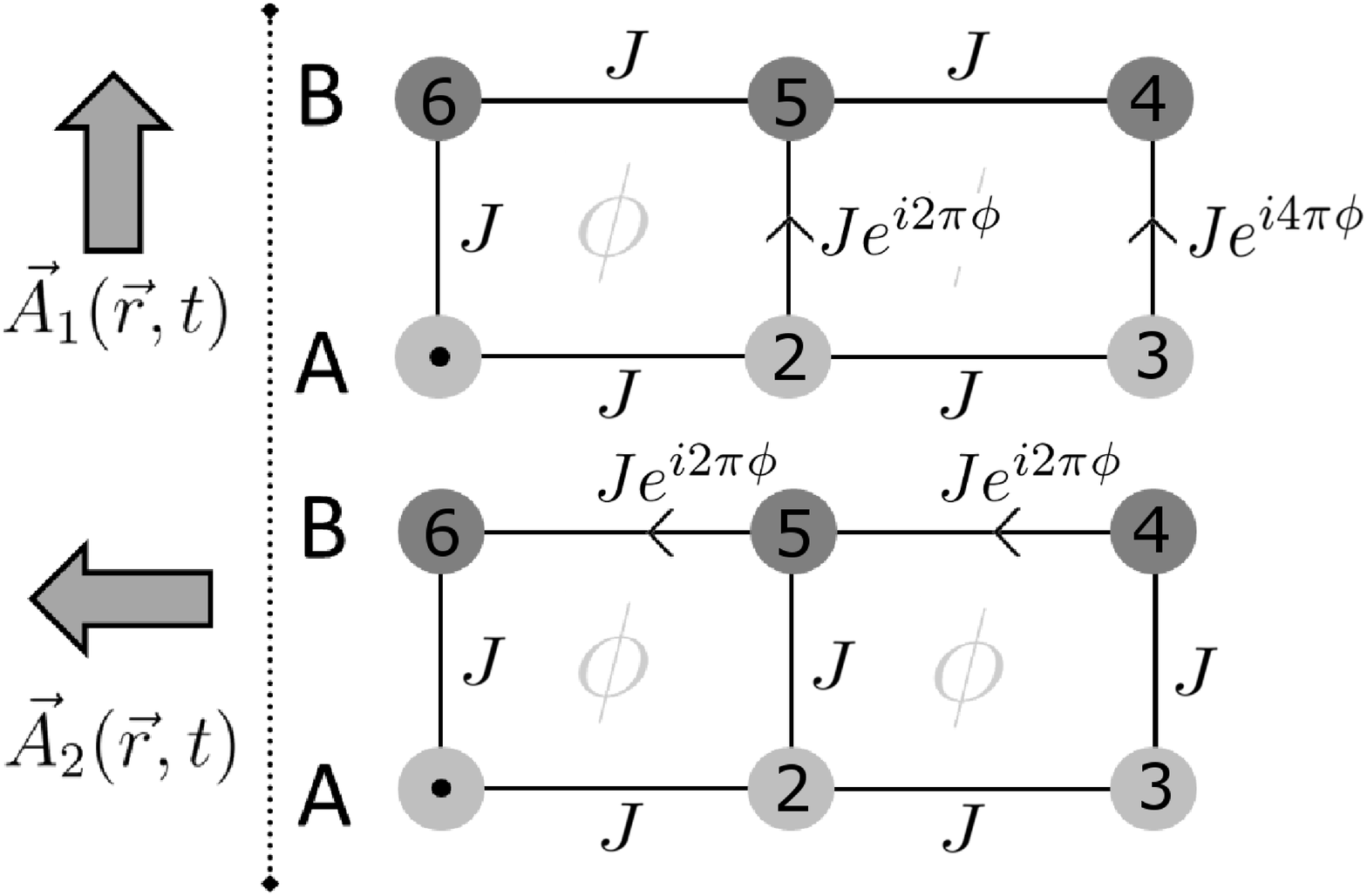}
\caption{The minimal model, a six site tight-binding lattice formed by two adjacent squares. Two different gauge choices $\vec{A}_1(\vec{r},t)$, $\Phi_1(\vec{r},t)$ and $\vec{A}_2(\vec{r},t)$, $\Phi_2(\vec{r},t)$ create Hamiltonians (Eq.\ref{H1} and Eq.\ref{H2}) with Peierls phases as shown. Note that the origin is chosen as the bottom left site.}
\label{fig:twoSquares}
\end{figure}

In this section, we study a model which can be used to probe the validity of the above interpretations, particularly for time dependent situations, in a simple way. We consider a six site tight-binding model, formed by two adjacent squares as shown in Fig.\ref{fig:twoSquares}. Such an arrangement can be experimentally realized by isolating two sites of the optical lattice in a spin-1 synthetic dimension\cite{GoldmanArtGaugeFieldInSyntheDim}. The hopping between neighboring sites is taken as $J$, the left bottom site is the origin, long side of the rectangle is taken as the x-axis. We simulate the situation where the fluxes through both squares are linearly ramped from zero to a final value $\phi$ (in units of flux quantum $h/e$), in a time scale $\tau$. The magnetic field in this system is $\vec{B}(t)= B_0 f_\tau(t) \hat{z}$, where the ramp function is
\begin{equation}
f_\tau(t) = \left\{ \begin{array}{r@{\quad:\quad}l} 0 & t \leq 0 \\ \frac{t}{\tau} & 0<t<\tau \\ 1 & t \geq \tau \end{array} \right.
\end{equation}
and $\phi= \frac{B_0 a^2}{h/e}$ for lattice constant $a$.

One set of potentials that create this time dependent flux is
\begin{eqnarray}
  \vec{A}_1(\vec{r},t) &=& f_\tau(t) B_0 x \hat{y}, \\
  \Phi_1(\vec{r},t) &=& 0.
\end{eqnarray}
which creates the following lattice Hamiltonian
\begin{equation}\label{H1}
H_1(t)= -J \left( a_1^\dagger a_2 + a_2^\dagger a_3 + a_4^\dagger a_5 +a_5^\dagger a_6 + a_6^\dagger a_1 + h.c. \right)
- J \left( e^{-i 2 \pi \phi f_\tau(t)} a_2^\dagger a_5 + e^{-i 4 \pi \phi f_\tau(t)} a_3^\dagger a_4 + h.c. \right),
\end{equation}
utilizing the Peierls substitution $J \to J \exp{\big(- i e/\hbar \int_{\vec{R}_i}^{\vec{R}_f} \vec{A}\cdot d l \big)}$. The line integral is evaluated along the classical path from the initial to the final lattice points.

Same time dependent flux can also be generated by
\begin{eqnarray}
  \vec{A}_2(\vec{r},t) &=&  - f_\tau(t) B_0 y \hat{x}, \\
  \Phi_2(\vec{r},t) &=& 0.
\end{eqnarray}
The corresponding lattice Hamiltonian is
\begin{equation}\label{H2}
H_2(t)= -J \left( a_1^\dagger a_2 + a_2^\dagger a_3 + a_2^\dagger a_5 +a_3^\dagger a_4 + a_6^\dagger a_1 + h.c. \right)
- J \left( e^{-i 2 \pi \phi f_\tau(t)} a_4^\dagger a_5 + e^{-i 2 \pi \phi f_\tau(t)} a_5^\dagger a_6 + h.c. \right).
\end{equation}
When the flux is static,i.e. $t<0$ and $t>\tau$, these two Hamiltonians are related by a gauge transformation. However, they are not equivalent during the ramp up interval, as can be seen by considering a third set of potentials.
\begin{eqnarray}
  \vec{A}_3(\vec{r},t) &=&  - f_\tau(t) B_0 y \hat{x}, \\
  \Phi_3(\vec{r},t) &=& \frac{B_0}{\tau} x y \left(\theta(t)-\theta(t-\tau) \right),
\end{eqnarray}
where $\theta$ is the step function. The lattice Hamiltonian for this set of potentials has on site terms
\begin{equation}\label{H3}
H_3(t)=H_2(t)+ J \left[  2 \pi \phi \gamma a_5^\dagger a_5 +  4 \pi \phi \gamma a_4^\dagger a_4 \right] \left(\theta(t)-\theta(t-\tau) \right).
\end{equation}
Here $\gamma=\frac{\hbar}{J \tau}$ is a dimensionless parameter measuring the adiabaticity of the ramp up, with $\gamma=0$ corresponding to the adiabatic limit.

\begin{figure}
\includegraphics[scale=0.55]{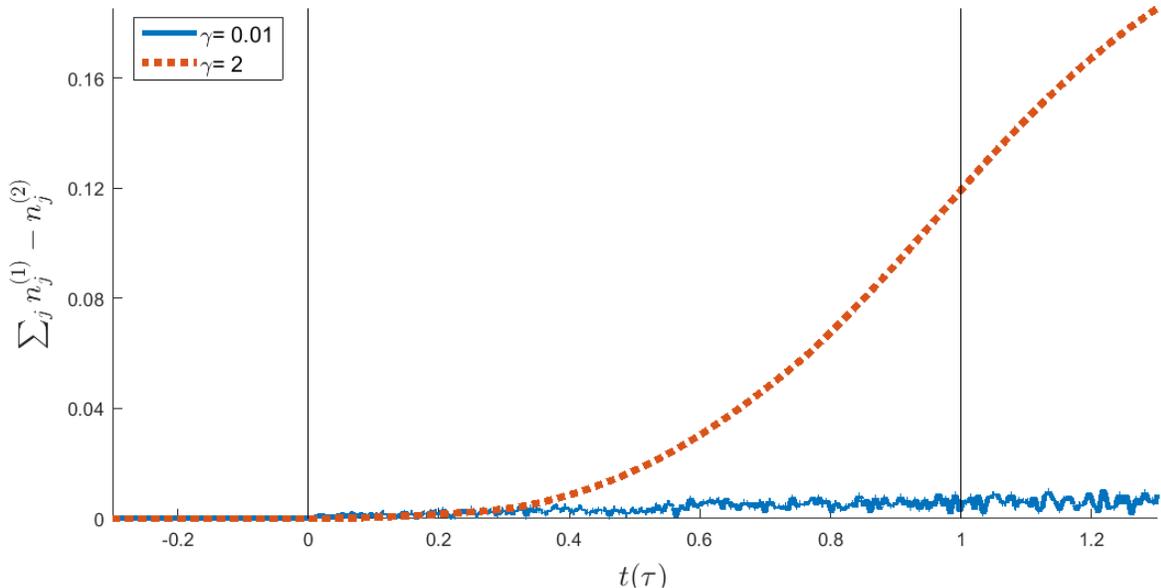}
\caption{The total difference between the site densities of the wavefunctions evolved with $H_1$ and $H_2$ as a function of time. The magnetic flux is ramped up from $\phi = 0$ to $\phi = 1/3$ between  $0 \le t \le \tau$ with two different ramping parameters, $\gamma = \{0.01, 2\}$. For $\gamma = 0.01$, the ramp of magnetic flux is slow and adiabatic, the difference in total site densities is less than $0.01\%$ which is within our numerical accuracy. For $\gamma = 2$, the ramping is non-adiabatic and results in distinct time evolution for $|\psi_1|^2$ and $|\psi_2|^2$.}
\label{fig:errorTrack}
\end{figure}

To investigate the gauge dependence, we prepared same initial site distribution under zero magnetic field. Then, each system is numerically time evolved, where the onset and the full turn-on of the flux occur at $t=0$ and $t=\tau$. The time evolution is done by Cayley's form, with the time evolution operator $\mathcal{T}\exp{(-\frac{i}{\hbar} \int^{t+\Delta t}_{t} \hat{H}(\tau) d \tau )} \approx (1 - i H \Delta t /2)/(1 + i H \Delta t/2)$. This form provides an accurate evolution of the wavefunction as long as the time step is much smaller than the typical time scale for the system.
We compare the density at each site $\langle a_i^\dagger a_i \rangle$ as a function of normalized dimensionless time. Note that all three Hamiltonians are related by a static gauge transformation for $t\le0$ and $t\ge \tau$ whereas only the first and the third Hamiltonian are related by a dynamical gauge transform also in the interval $0 \le t \le \tau$.

\begin{figure}
\includegraphics[width=0.95\textwidth]{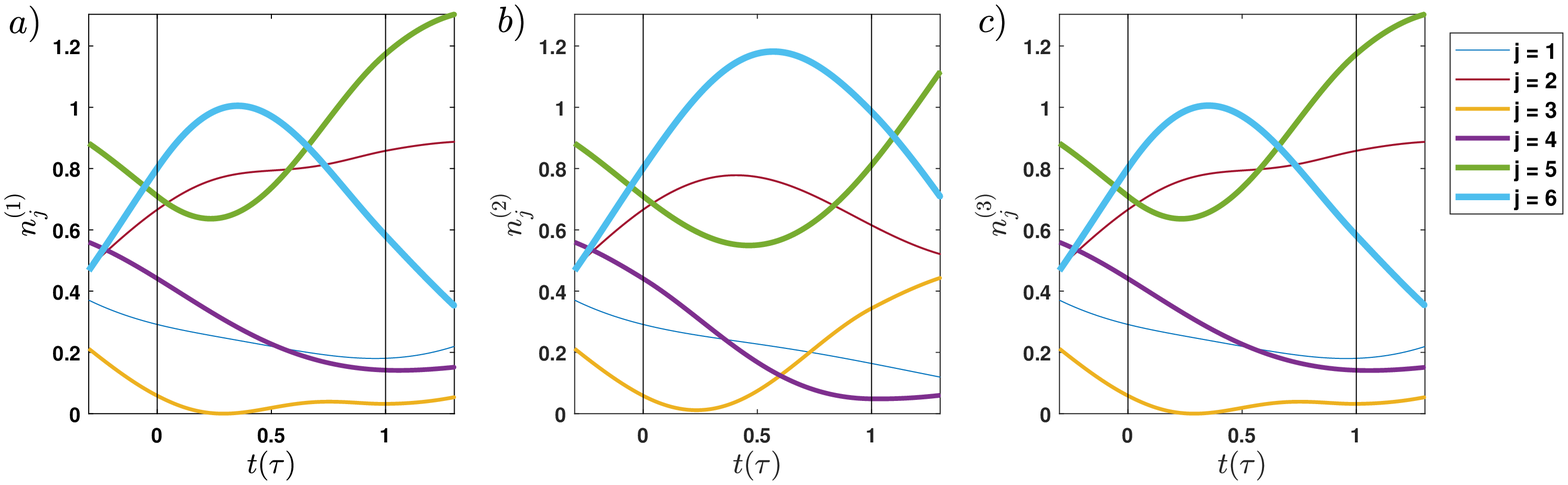}
\caption{(Color online) Minimal model time evolution of each site density $\langle a^{\dagger}_i a_i \rangle$ in the non-adiabatic region, $\gamma = 2$. The magnetic flux is ramped from $0$ to $1/3$ and the same initial wavefunction is time evolved for three Hamiltonians, $H_1$, $H_2$ and $H_3$. $H_1$ (Eq.\ref{H1}) and $H_3$ (Eq.\ref{H3}) are connected by a dynamic gauge transformation and give identical time-evolution for an arbitrary ramping rate, as shown in plot $a$ and $c$. Whereas the second Hamiltonian, $H_2$ (Eq.\ref{H2}) lacks the induced scalar potential and the density of each site in plot $b$ are clearly different from two other gauges.}
\label{fig:minimalEvolution}
\end{figure}

For close to adiabatic evolution $\gamma \ll 1 $ , the difference between total density of each site with reference to results of $H_1$ are negligibly small (see Fig.\ref{fig:errorTrack}). This is expected because at each moment of time the first two Hamiltonians can be related by a unitary transformation $U(t)^\dagger H_2(t) U(t) =H_1(t)$, and adiabatic evolution means that time dependence of that unitary transformation is negligible. The on site terms in the third Hamiltonian are zero in the truly adiabatic limit as they scale with $\gamma$. For $\gamma \gtrsim 1$ the evolution is non-adiabatic, and the results of $H_2$ are clearly distinct from $H_1$ and $H_3$. (See Fig.\ref{fig:minimalEvolution} for results with $\gamma = 2$). As the terms generated by the time derivative of the unitary transformation relating $H_1$ and $H_2$ are no longer negligible, the equivalence between the Hamiltonian based on {\bf static} gauge invariance is not correct any more. In the third Hamiltonian, these terms are cancelled exactly by the added on site potentials, and the equivalence between the first and the third Hamiltonians is a result of {\bf dynamical} gauge invariance (see Fig.\ref{fig:minimalEvolution}).

In a sudden quench, with $\gamma \to \infty$, the first and second Hamiltonians are related by a static gauge transformation, at all times except at $t=0$. The results of evolution after the quench clearly show that the evolution of site densities are different (See Fig.\ref{fig:minimalEvolution}) for $\gamma = 2$. This difference, which can be called static gauge dependence, is because the two Hamiltonians not connected by a gauge transformation exactly at the moment of the quench. The third Hamiltonian which is related to the first by a proper dynamical gauge transformation, has on site terms which act for an infinitesimal amount of time $\tau$, but their strength is as high as $\gamma \sim 1/\tau$. As a result these on site terms can generate the necessary phases creating the gauge independence. Hence, in a quench experiment, as long as the Hamiltonians are related only by a static gauge transformation, one can interpret the results to be gauge dependent because gauge equivalence is broken at the moment of quench. Throughout the rest of the paper, we will mention gauge dependent results in this sense.

Another, equally valid, interpretation is to say that a quench with first Hamiltonian and with second Hamiltonian correspond to two different physical scenarios. While both have the same magnetic field jumping from zero to $B_0$, the generated electric fields in the two cases are different. The difference between the momentary artificial electric fields is clearly seen as the delta function on-site potentials in the third Hamiltonian. In a cold atom experiment such (delta function in time) on-site potentials would be extremely hard to generate. Thus, quenches in experiments will in general be gauge dependent in the static gauge equivalence sense as discussed above.

\section{Two-Leg and Three-Leg Ladders}

In this section, we investigate the dynamics after the artificial magnetic field quench for two leg and three leg infinite ladders. The two leg system with a magnetic flux through each plaquette has been realized in real space\citep{Bloch2Leg}. Although it seems quite hard for a sudden artificial magnetic field change to be implemented in that particular configuration, the two leg system is simple enough to allow for analytical results amenable to interpretation. The three leg ladder is formed as a synthetic dimension experiment\cite{GoldmanArtGaugeFieldInSyntheDim}, by treating the hyperfine $F=1$ manifold as the synthetic dimension. Our investigation will focus on the dynamics of a single wavepacket which is an eigenstate before the quench, as this method is used commonly with Bosonic systems. A similar analysis can be carried out for Fermions by a suitable sum over the Brillouin zone.

\begin{figure}
\includegraphics[width=0.9 \textwidth]{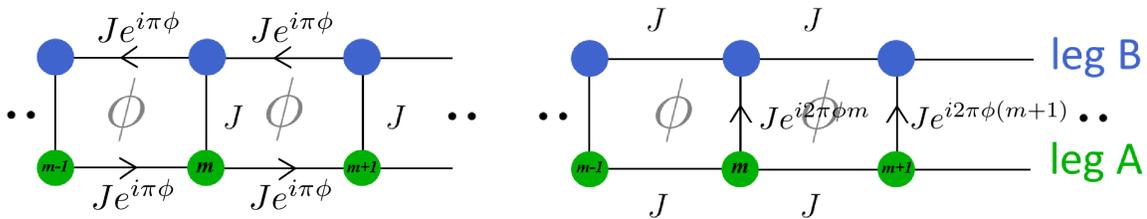}
\caption{(Color online) The schematic representation of the two-leg tight-binding ladder for two different gauge choices in Eq.\ref{Ham1} and Eq.\ref{Ham2}.}
\label{fig:twoLegModel}
\end{figure}
We first start with the two leg ladder, where the gauge is chosen so that hopping phases are equal and opposite on the two legs (see Fig.\ref{fig:twoLegModel})
\begin{equation}\label{Ham1}
\hat{H}^{(1)}_{2L} = - J \sum_{m = -\infty}^{\infty} \big( e^{i \pi \phi} a^{\dagger}_{m+1} a_{m} + e^{-i \pi \phi} b^{\dagger}_{m+1} b_{m} + b^{\dagger}_m a_{m} + H.c. \big)
\end{equation}
Defining Fourier transformed operators
\begin{equation}
\Big(a^{\dagger}_k,b^{\dagger}_k\Big) =\frac{1}{\sqrt{N}} \sum_{m=1}^{N} e^{i k m} \Big(a^{\dagger}_m,b^{\dagger}_m\Big),
\end{equation}
reduces the Hamiltonian to the $2 \times 2$ matrix at each k point.
\begin{equation}
\hat{H}_{2L}^{(1)}(k) = - 2 J \cos{(k)} \cos{(\pi \phi)} \hat{I} + \hat{h}(k)\cdot \mathbf{\sigma},
\end{equation}
where
$\hat{h}(k)\cdot \mathbf{\sigma}/J = - 2 \sin{(k)} \sin{(\pi \phi)} \hat{\sigma}_z - \hat{\sigma}_x$. Diagonalizing the matrix results in two bands, for the eigenvalue equation,
\begin{equation}\label{EigEqn2}
\hat{H}_{2L}^{(1)}(k) |k,n;\phi\rangle = E |k,n;\phi \rangle ,
\end{equation}
and $n \in \{1,2\}$ yields energies
\begin{equation}
\frac{E^{1,2}(k)}{J} = - 2 \cos{k} \cos{(\pi \phi)} \pm \sqrt{4 \sin^2{k} \sin^2{(\pi \phi)} + 1},
\end{equation}
and the corresponding creation operators are
$$\begin{pmatrix} \hat{\alpha}_k \\ \hat{\beta}_k\end{pmatrix} =  \begin{pmatrix} \cos \theta^{\phi}_k/2 & \sin \theta^{\phi}_k/2\\ -\sin \theta^{\phi}_k/2 & \cos \theta^{\phi}_k/2 \end{pmatrix} \begin{pmatrix} \hat{a}_k \\ \hat{b}_k\end{pmatrix},$$ with $\tan \theta^{\phi}_k = \frac{1}{2 \sin k \sin{(\pi \phi)}}$.

We now consider an initial state which is a wavepacket made up of a superposition of states in one band near quasimomentum $k_0$, at an initial flux value $\phi_1$. We then abruptly change the value of the flux to $\phi_2$ and investigate the dynamics of the wavepacket.
If the pre-quench wavefunction is a wavepacket of the lower band, its center of mass will be moving with the group velocity $v_g \propto \partial E^{1} / \partial k$. After the quench this packet will be a superposition of the new upper and lower band states, which will in general have different group velocities. As the system evolves, the components in the upper and lower bands will separate into two wavepackets in a time scale determined by the spatial extent of the original wavepacket divided by the group velocity difference between the upper and lower bands in the new Hamiltonian. The density in each packet will be proportional to the overlaps of before quench and after quench bands $C^{\phi_1,\phi_2}_{n_1,n_2} (k) = \langle k, n_1; \phi_1 | k, n_2; \phi_2 \rangle$, where $n_1,n_2 \in \{1,2 \}$ are the band indices. Hence, any in situ method which probes density can directly measure these coefficients. These coefficients are found as
\begin{eqnarray}\label{overlapCoeff}
C^{\phi_1,\phi_2} (k) &=& \begin{pmatrix} \cos \Big( \frac{\theta^{\phi_2}_{k} - \theta^{\phi_1}_{k}}{2} \Big) & \sin \Big( \frac{\theta^{\phi_2}_{k} - \theta^{\phi_1}_{k}}{2} \Big)\\ -\sin \Big( \frac{\theta^{\phi_2}_{k} - \theta^{\phi_1}_{k}}{2} \Big) & \cos \Big( \frac{\theta^{\phi_2}_{k} - \theta^{\phi_1}_{k}}{2} \Big) \end{pmatrix}.\\
\cos \Big( \frac{\theta^{\phi_2}_{k} - \theta^{\phi_1}_{k}}{2} \Big) &=& \frac{1}{2}\Big( \sqrt{1 + \frac{b_2}{h_2}}\sqrt{1 + \frac{b_1}{h_1}}+ \sqrt{1 - \frac{b_2}{h_2}}\sqrt{1 - \frac{b_1}{h_1}} \Big),\\
\sin \Big( \frac{\theta^{\phi_2}_{k} - \theta^{\phi_1}_{k}}{2} \Big) &=& \frac{1}{2}\Big( \sqrt{1 - \frac{b_2}{h_2}}\sqrt{1 + \frac{b_1}{h_1}}- \sqrt{1 + \frac{b_2}{h_2}}\sqrt{1 - \frac{b_1}{h_1}} \Big),\\
\end{eqnarray}
where $b_i = -2 \sin(k) \sin(\pi \phi_i)$ and $h_i = \sqrt{4 \sin^2{k} \sin^2{(\pi \phi_i)} + 1}$.

For example, $|C^{\phi_1,\phi_2}_{1,1}|^2$ is the probability that the initial state in the lower band of $\phi_1$ to stay in the new lower band of $\phi_2$ etc. Let's assume that at t=0 we have the Hamiltonian with flux $\phi_1$, and a particle is in an eigenstate at quasimomentum k, and band index $1$ or $2$. A sudden change in the flux to $\phi_2$ will not immediately change the wavefunction, however its subsequent evolution will be governed by the new Hamiltonian at $\phi_2$. As both Hamiltonians before and after the quench have discrete translation symmetry along the ladder, quasimomentum will be a conserved quantity.

We simulate such a scenario using a Gaussian wavefunction (Eq.\ref{initPsi}) with a width of sixty lattice sites along $\hat{x}$-axis, $\Delta = 60$, and the total system has $N = 1200$ sites in each leg. This function is localized in the lowest band at $k=k_0$ in momentum space and centred in the middle of finite spatial coordinates $\phi_1=0$,
\begin{equation}\label{initPsi}
\psi_0(m)= \frac{1}{(\pi \Delta^2)^{1/4}} e^{-\frac{(m- m_0)^2}{2 \Delta^2}} e^{i k_0 m}.
\end{equation}

We evolve the wavefunction in Eq.\ref{initPsi} with $k_0 = \pi/4$ in the quenched Hamiltonian $\phi_2=1/6$. As the group velocities of the upper and lower bands are slightly different from the pre-quenched bands for small values of difference in $\phi$, the wavepacket splits into two packets going in the same direction. The total weight of the two packets from the simulation $P_1= 0.91$ and $P_2=0.09$ are in excellent agreement with the overlap coefficients $|C^{0,1/6}_{11}|^2 = 0.91$ and $|C^{0,1/6}_{12}|^2 = 0.09$ with an error less than $\le \% 0.2$.
\begin{figure}
\includegraphics[width=1\textwidth,height=0.25\textwidth] {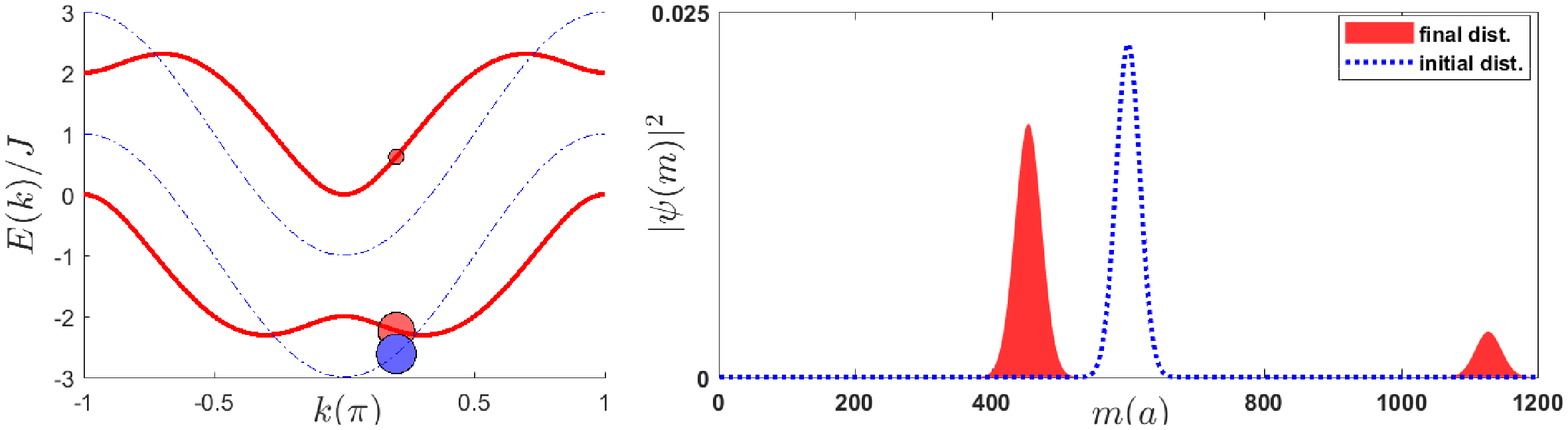}
\includegraphics[width=1\textwidth,height=0.25\textwidth] {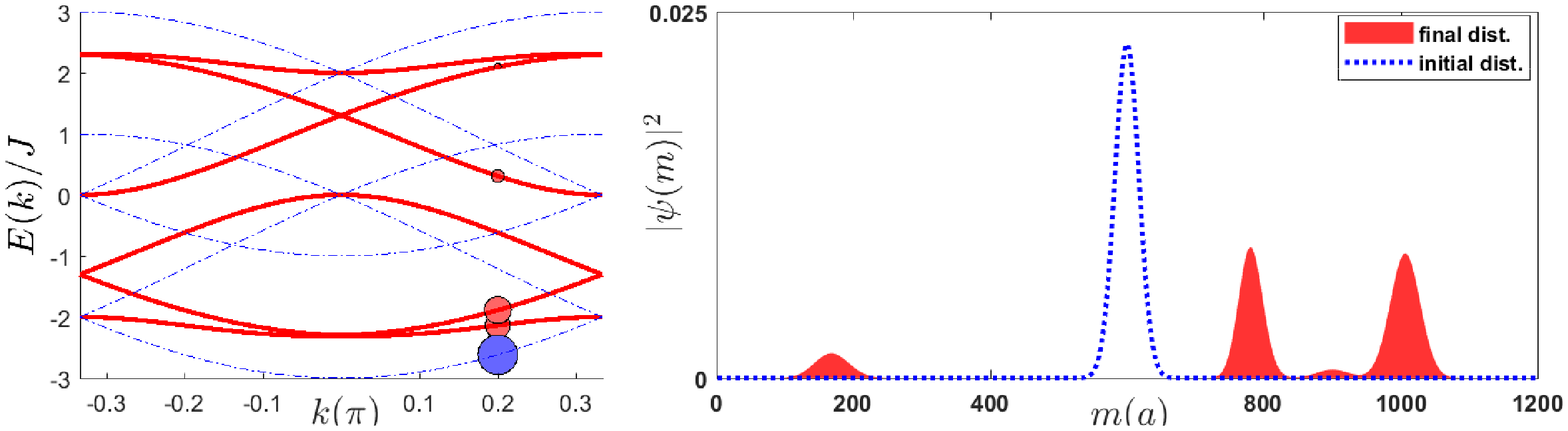}
\caption{(Color Online) The splitting of the same initial wavepacket after the quench for the two gauge choices in Eq.\ref{Ham1} and Eq.\ref{Ham2}. Initial wavepacket at zero magnetic flux prepared at $k = 0.2 \pi$ and the artificial magnetic flux is quenched as $\phi = 0 \to 1/3$. In the first gauge choice (upper plot), wavepacket splits into two packets moving in opposite directions while it splits into four packets in the second gauge choice (lower plot). The pre- (dashed lines) and the post-quench (solid lines) band structures are shown on the left plots. The initial and the final weights of the packet in each band are shown with blue (dark grey) and  red (grey) filled circles on the energy bands.}
\label{fig:twoLegBandSplitting}
\end{figure}

The separation of the wavepacket can be dramatic if the group velocities of the two bands have opposite signs. A right moving wavepacket at $k=\pi/5$ at $\phi_1=0$ splits into a left moving wavepacket and right moving wavepacket, as seen in the first plot of Fig.\ref{fig:twoLegBandSplitting}. The weights of the two packets once again follow the simple overlap formulas, and for this case the wavepacket which reverses the direction is considerably more prominent. While quasimomentum is a conserved quantity during the quench, the same can not be said of momentum. This can either be interpreted as the effect of the change in the definition of canonical momentum, or the effect of the induced electric field during the quench, as discussed in the previous section.
\begin{figure}
\includegraphics[width=0.55 \textwidth]{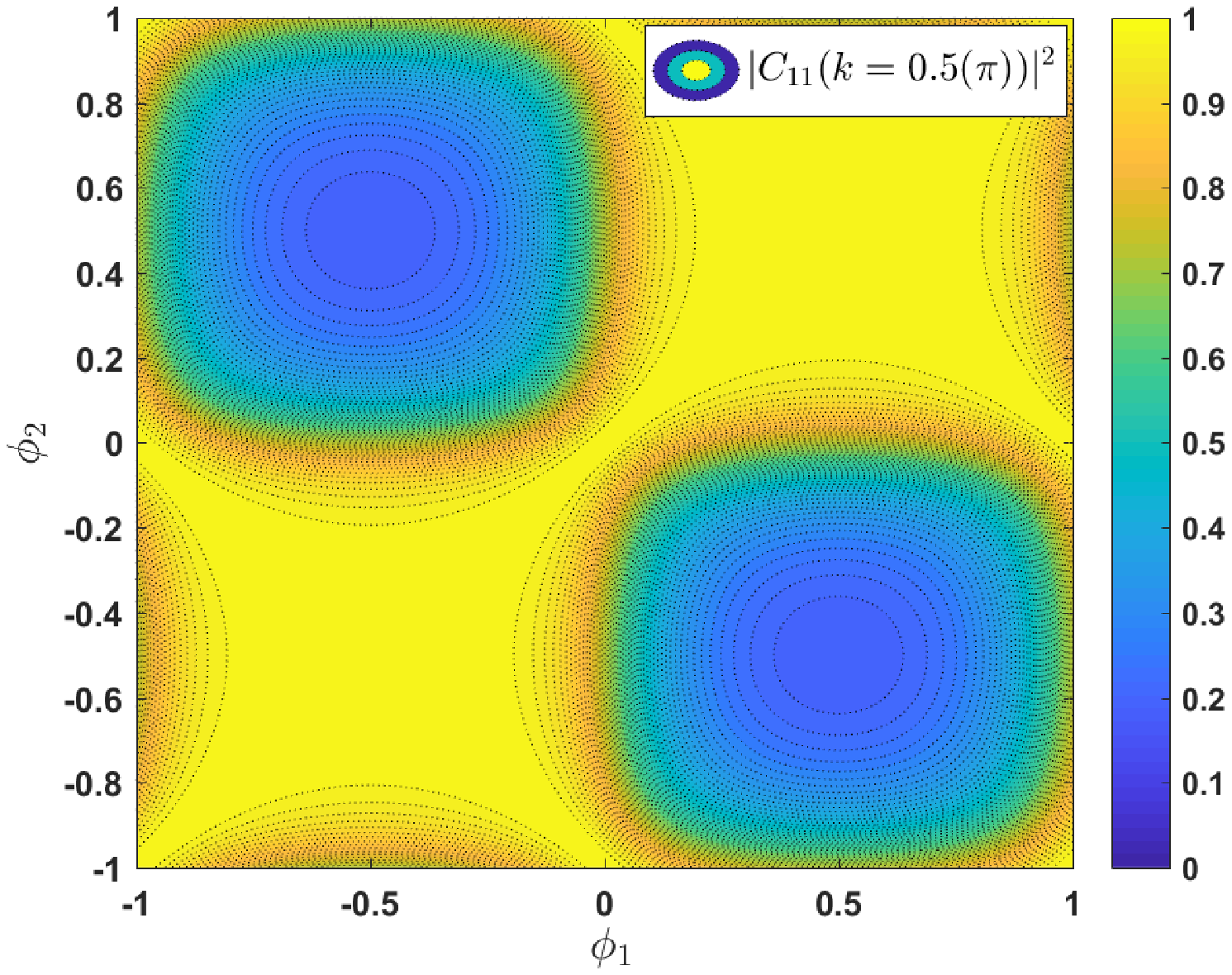}
\caption{(Color online) The probability of the initial wavepacket prepared in the lowest band to maintain its band, which is calculated in Eq.\ref{overlapCoeff} as $|C^{\phi_1,\phi_2}_{11}(k)|^2$ after the quench between two arbitrary magnetic fluxes, $\phi_1$ and $\phi_2$. The initial packet is prepared around $k=\pi/2$. Note that $|C^{\phi_1,\phi_2}_{11}(k)|^2 + |C^{\phi_1,\phi_2}_{12}(k)|^2 = 1$ and $|C^{\phi_1,\phi_2}_{11}(k)|^2 = |C^{\phi_1,\phi_2}_{22}(k)|^2$.}
\label{fig:overlapCoeffs}
\end{figure}

In Fig.\ref{fig:overlapCoeffs}, we calculate the distribution of a wavepacket prepared in the lowest band of Hamiltonian with initial magnetic flux $\phi_1$ at $k=\pi/2$ onto the eigenstates of the arbitrary final flux $\phi_2$ as calculated in Eq.\ref{overlapCoeff}. The coefficients, $|C^{\phi_1,\phi_2}_{11}(k)|^2$ and $|C^{\phi_1,\phi_2}_{12}(k)|^2$  determine the weights of the two wavepackets after the quench. The packet transfers significant weight into the other band only if the initial or the final magnetic flux is close to zero or one. The initial packet mostly preserves its band index unless the direction of the magnetic field is reversed.

While the dynamics of the wavepacket in a quench made with $H_{2L}^{(1)}$ is easy to understand, this Hamiltonian is not unique in describing a magnetic flux quench.
We investigate the effect of the gauge choice by considering the two leg ladder for which the Peierls phases are defined on the rungs of the ladder, rather than on the legs (see second plot in Fig.\ref{fig:twoLegModel})
\begin{equation}\label{Ham2}
\hat{H}^{(2)}_{2L} = - J \sum_{m = -\infty}^{\infty} \big( a^{\dagger}_{m+1} a_{m} + b^{\dagger}_{m+1} b_{m} + e^{i 2 \pi \phi m} b^{\dagger}_m a_{m} + H.c. \big)
\end{equation}

The most important difference between two Hamiltonians for the gauges is that $H^{(1)}_{2L}$ preserves the full translational invariance of the ladder, while the second choice $H^{(2)}_{2L}$ explicitly breaks it. For a rational value of flux $\phi=p/q$, the first Hamiltonian has two bands, defined throughout the first Brillouin zone $-\pi<k<\pi$, while the second has $2 q$ bands, defined in the magnetic Brillouin zone, $-\pi/q<k<\pi/q$. As we have seen in the previous section that measurable quantities depend on the choice of gauge, it is necessary to ask whether the number of packets which an initial Gaussian packet splits into can be changed. If the initial wavepacket is localized in a single band around an eigenstate of the two leg ladder with zero flux, it is quite clear that it will split into two upon a quench to $H^{(1)}_{2L}$. In such a quench, the translational symmetry along the ladder is not broken, and the final Hamiltonian has only two bands at a given quasimomentum. If the number of the final wavepackets were only controlled by the number of bands in the final Hamiltonian, a quench to $H^{(2)}_{2L}$ would be expected to generate $2 q$ wavepackets. Such an expectation is not reasonable, as it would predict an infinite number of wavepackets for an irrational $\phi$.

The number of wavepackets after the quench is not controlled by the number of bands in the final Hamiltonian, as the way the translational symmetry is broken in $H^{(2)}_{2L}$ is not arbitrary. Irrespective of the value of the flux, $H^{(2)}_{2L}$ is unitarily connected to a translational symmetric system $H^{(1)}_{2L}$
\begin{equation}
H^{(2)}_{2L}= U H^{(1)}_{2L} U^{\dagger},
\end{equation}
which implies a relation between the bands of $H^{(2)}_{2L}$. The unitary operator connecting the two static gauge choices is
\begin{equation}\label{gauge2tran}
U= e^{i \pi \phi \sum_{m} m(b^{\dagger}_m b_m - a^{\dagger}_m a_m)}.
\end{equation}
If we start with an eigenstate of the ladder at zero field $|\psi_{k_0}(0)\rangle$, and investigate the evolution after quenches to $H^{(1)}_{2L}$ and $H^{(2)}_{2L}$, the resulting state at time $t$,
\begin{eqnarray}\label{timEvo2Leg}
|\psi_{k_0}(t)\rangle_1 &=& e^{-i H^{(1)}_{2L} t/\hbar} |\psi_{k_0}(0)\rangle. \nonumber \\
|\psi_{k_0}(t)\rangle_2 &=& e^{-i H^{(2)}_{2L} t/\hbar} |\psi_{k_0}(0)\rangle, \nonumber \\
&=& U e^{-i H^{(1)}_{2L} t/\hbar} U^\dagger |\psi_{k_0}(0)\rangle.
\end{eqnarray}
We obtain the last expression by using the static gauge equivalence in Eq.\ref{gauge2tran}. For the case of $|\psi_k(t)\rangle_1$, we have two components which correspond to the two bands which are at the same $k$ vector.

We need to calculate how many components the same initial wavefunction will develop if the time evolution is done with $H^{(2)}_{2L}$. The unitary operator in Eq.\ref{timEvo2Leg}, $U^{\dagger}$ attaches a different phase to each lattice site which does not change the density in real space, hence it can not change the number of components after a flux quench. The remaining part excluding the time evolution operator of $H^{(1)}_{2L}$, the state $U^\dagger |\psi_{k_0}(0)\rangle$ can be expanded as the eigenstates of $H^{(1)}_{2L}$. Without a loss of generality (see Appendix\ref{Appnd}), assume that the initial state is in the lowest band of the zero field ladder at lattice momentum $k_0$,  $|\psi_{k_0}(0)\rangle = |k_0,1;\phi = 0\rangle$. Then,
\begin{eqnarray}
U^\dagger |k,1;\phi = 0 \rangle = &+&\frac{1}{\sqrt{2}} \cos{\frac{\theta^{\phi}_{k+\pi \phi}}{2}} | k+ \pi \phi  ,1;\phi \rangle - \frac{1}{\sqrt{2}} \sin{\frac{\theta^{\phi}_{k+\pi \phi}}{2}} | k+ \pi \phi  ,2;\phi \rangle  \nonumber \\
&-& \frac{1}{\sqrt{2}} \sin{\frac{\theta^{\phi}_{k-\pi \phi}}{2} }| k- \pi \phi  ,1;\phi \rangle - \frac{1}{\sqrt{2}} \cos{\frac{\theta^{\phi}_{k-\pi \phi}}{2}} | k- \pi \phi  ,2;\phi \rangle.
\end{eqnarray}

The effect of $U^\dagger$ on the wavepacket is to displace the wavefunction components in the upper and lower legs by equal and opposite momentum shifts. Hence this packet has two packets in each band of two momentum components at $k+\pi \phi$ and $k-\pi \phi$, and quench to $H^{(2)}_{2L}$ can generate at most four split packets after the quench. As an example consider a quench of the initial state as seen in Fig.\ref{fig:twoLegBandSplitting}, with $k= 0.2 \pi$, it generates two wavepackets after evolution with $H^{(1)}_{2L}$ and four wavepackets after evolution with $H^{(2)}_{2L}$.

While two leg ladders can be realized with real space optical lattices\cite{Bloch2Leg} a sudden change in the artificial magnetic fluxes generated by methods modifying the real space hopping between sites would not be easy to implement. For three leg ladders, real space implementation would have to become even more involved. However, Raman coupling between the hyperfine levels of the atoms which constitutes hopping in the synthetic dimension provides the means to not only implement wider ladders, but to modify the effective gauge fields at a fast time scale. We study the effect of an artificial magnetic field quench for a three leg ladder as can be obtained by using the spin-1 hyperfine manifold as the synthetic dimension.

\begin{figure}
\centerline{\includegraphics[width=1.2\textwidth]{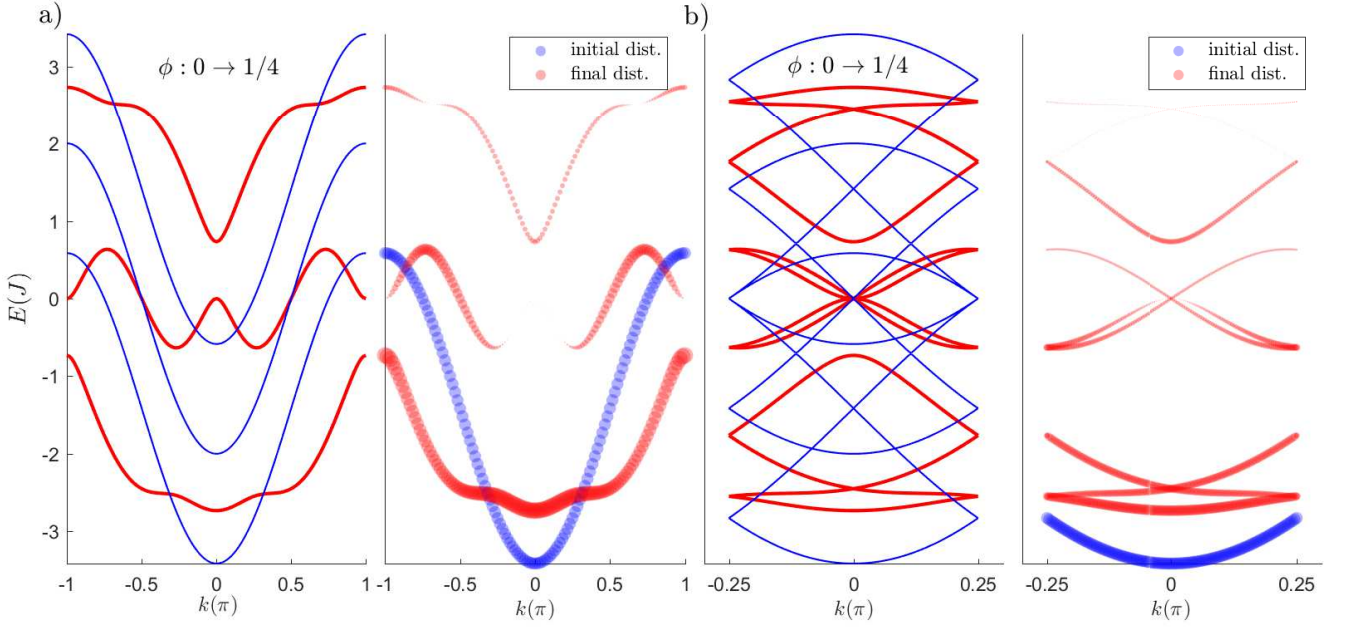}}
\caption{(Color online) The pre- (dashed lines) and post-quench (solid lines) band structures of the three-leg ladder for two gauge choices are plotted in plot $a$ and $b$. The magnetic flux is quenched as $\phi = 0 \to 1/4$. The plot $c$ ($d$) shows the weight of the initial packet prepared in the lowest band of zero field Hamiltonian on the new bands at $\phi = 1/4$. The blue (dark grey) line is the weight of the initial packet at each momentum, and the red lines with varying thickness give the weights after the quench for the gauge choice 1 (2). In the second gauge choice, the initial packet at each momentum splits into nine bands in the reduced Brillouin zone. The static gauge transformation connecting the two Hamiltonians puts a constraint on the maximum number of packets that an initial packet can split into, the square of the number of legs $3\times3 =9$.}
\label{fig:3LegBandDistribution}
\end{figure}

As in the two leg case, we choose two Hamiltonians for three leg ladder both with flux $\phi$ through every plaquette albeit with different gauge choices. For the first gauge, the phases are added to the hopping along the real dimension, in the first and third legs:
\begin{equation}\label{Ham3leg1}
\hat{H}^{(1)}_{3L} = - J \sum_{m = -\infty}^{\infty} \big( e^{i 2 \pi \phi} a^{\dagger}_{m+1} a_m + b^{\dagger}_{m+1} b_m + e^{-i 2 \pi \phi} c^{\dagger}_{m+1} c_m + b^{\dagger}_m a_{m} + c^{\dagger}_m b_{m} +  H.c. \big),
\end{equation} while the second gauge choice modifies hopping along the synthetic dimension
\begin{equation}\label{Ham3leg2}
\hat{H}^{(2)}_{3L} = - J \sum_{m = -\infty}^{\infty} \big( a^{\dagger}_{m+1} a_m + b^{\dagger}_{m+1} b_m +  c^{\dagger}_{m+1} c_m + e^{i 2 \pi \phi m} b^{\dagger}_m a_{m} + e^{i 2 \pi \phi m} c^{\dagger}_m b_{m} +  H.c. \big).
\end{equation}
As these Hamiltonians are equivalent up to a static gauge transformation,
\begin{equation}\label{gauge3tran}
U = e^{i 2 \pi \phi \sum_m m (c_m^{\dagger} c_m - a_m^{\dagger} a_m }),
\end{equation}
and their energy spectra are the same, plotted with blue (dark grey) and red (light grey) colors for $\phi = 0$ and $\phi = 1/4$ in Fig.\ref{fig:3LegBandDistribution}.

We consider each of all initial states prepared at lowest band of the zero field spanning the Brillouin zone. We quench these initial states into two Hamiltonians in Eq.\ref{Ham3leg1} and Eq.\ref{Ham3leg2}, and calculate their distribution at each quasimomentum onto new bands (see Fig.\ref{fig:3LegBandDistribution}). The initial states are denoted with blue (dark grey) points and its distribution are denoted with red (grey) points. The size of red dots represents their weight in the quenched bands. For both gauge choices, notice that the energy bands are equivalent, where the energy bands in the second gauge are folded into the reduced Brillouin zone.

The dramatic change in the number of packets with the gauge choice cannot be simply explained by an increase in the number of bands upon folding to the reduced Brillouin zone. The unitary transformation between the two Hamiltonians restricts the final number of wavepackets acting similar to a selection rule. As discussed in the two leg ladder case, the Hamiltonian $\hat{H}^{(2)}_{3L}$ for $\phi_2=1/4$ is not periodic with a single site translation along the real dimension. Full symmetry is restored only after a translation by four lattice sites. Hence, as viewed in the reduced magnetic Brillouin zone, this Hamiltonian has not three but $3\times 4 = 12$ bands. In that respect, it may seem surprising that the quench results in the splitting of the wavepacket into at most nine packets instead of twelve. The connection to $\hat{H}^{(1)}_{3L}$ actually reflects the fact that a single site translation in real space for $\hat{H}^{(2)}_{3L}$ can be combined with a gauge transformation (Eq.\ref{gauge3tran}) to restore the symmetry. As a result, a quench to $\hat{H}^{(2)}_{3L}$ can result in at most $3\times 3=9$ wavepackets.

\begin{figure}
\includegraphics[width=1\textwidth,height=0.3\textwidth]{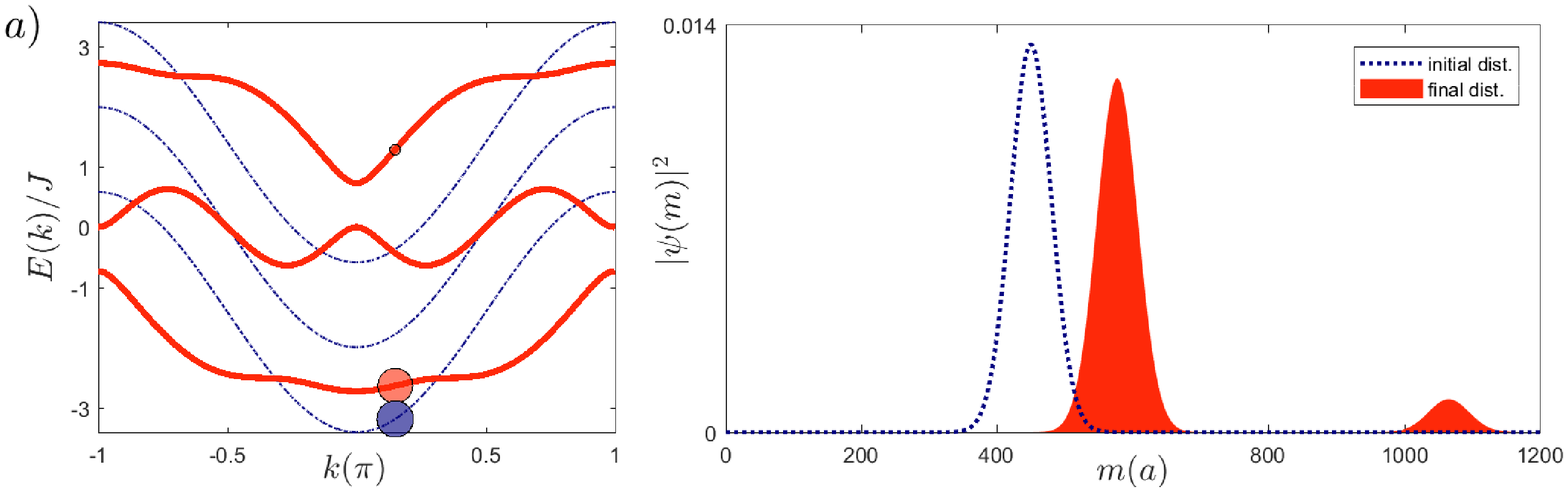}
\includegraphics[width=1\textwidth,height=0.3\textwidth]{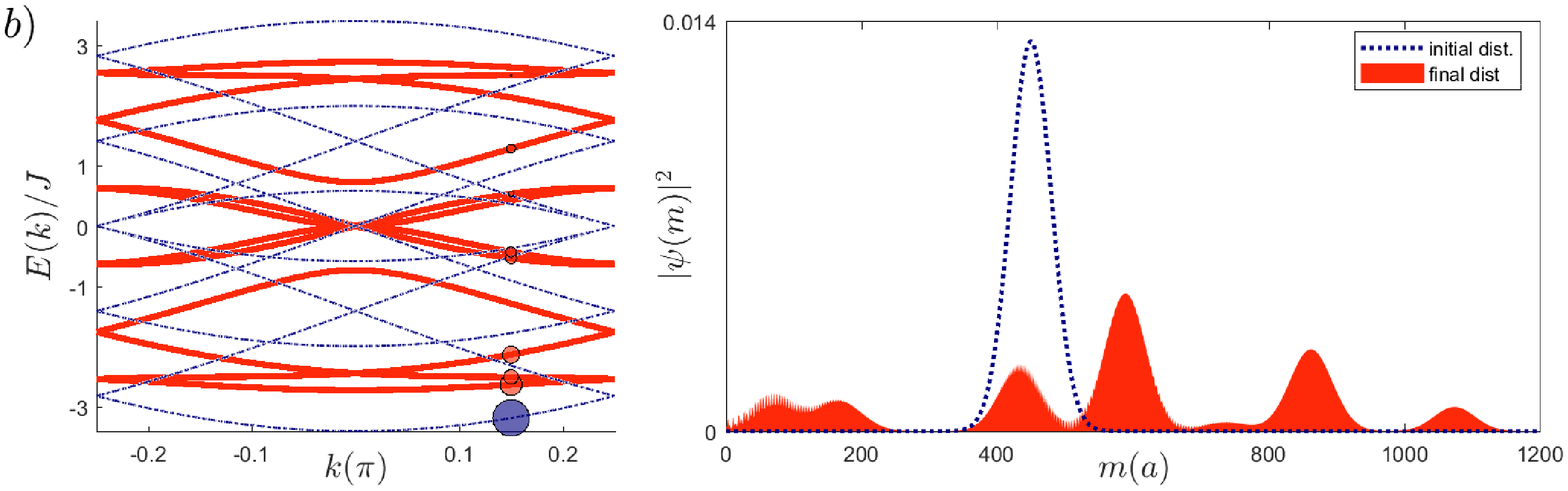}
\caption{(Color online) Time evolution of one wavepacket taken from Fig.\ref{fig:3LegBandDistribution}. The initial packet is prepared under zero field and quasimomentum at $k = 0.15 \pi$. Plot a) Flux quench $\phi = 0 \to 1/4$ in the first gauge choice results in two packets independent of the denominator of magnetic flux $q = 4$. Plot b) The same flux quench in the second gauge choice splits the initial packet into eight packets. The number of split packets are less than or equal to the square of the number of legs, $3 \times 3 = 9$, as expected.}
\label{fig:3LegPacketSplit}
\end{figure}

For example, we take one sample as an initial packet from Fig.\ref{fig:3LegBandDistribution} centred at a quasimomentum eigenstate for an initial flux $\phi_1=0$ which is narrow in momentum space. After the change in the flux to $\phi_2 = 1/4$, we expect the packet to split into at most three packets moving with different velocities as this is a system with three bands. Indeed, as can be seen in Fig.\ref{fig:3LegPacketSplit}, an initial wavefunction at $k=0.15 \pi$ of the lowest band of $\phi_1=0$ with a width of $60$ lattice sites is split into two wavepackets upon a quench into $\hat{H}^{(1)}_{3L}$, with $\phi=1/4$. The weight of each wavepacket is again determined by the overlap between the initial band and final bands at the same $k$ value. The same initial wavefunction splits into eight smaller wavepackets if the quench is carried out to $\hat{H}^{(2)}_{3L}$ (see Fig.\ref{fig:3LegPacketSplit}). The same argument is extended to a general N leg ladder in the appendix, showing that the maximum number of packets after the quench to a parallel gauge is $N^2$.

One of the most appealing features of synthetic dimension experiments is the ease with which hoppings along the synthetic dimension can be manipulated. In other words, among the two Hamiltonians considered above, $H^{(1)}_{3L}$ and $H^{(2)}_{3L}$ which are connected by a static gauge transformation, $H^{(2)}_{3L}$ is more readily implementable compared to $H^{(1)}_{3L}$ which has hopping phases along the real space dimension. In that regard, it is useful to consider scenarios where gauge dependence of the quench process can be explored through a manipulation of phases solely in the synthetic dimension. To this end, we introduce another Hamiltonian, which has zero flux per plaquette but has a built in phase angle $\Theta$ along the synthetic dimension.

\begin{equation}\label{Ham3legtheta}
\hat{H}^{\Theta}_{3L} = - J \sum_{m = -\infty}^{\infty} \big( a^{\dagger}_{m+1} a_m + b^{\dagger}_{m+1} b_m + c^{\dagger}_{m+1} c_m + e^{i 2 \pi \Theta} b^{\dagger}_m a_{m} + e^{i 2 \pi \Theta} c^{\dagger}_m b_{m} +  H.c. \big)
\end{equation}
Hamiltonians for different values of $\Theta$ are equivalent up to an almost trivial gauge transformation which is equivalent to a coordinate shift by $\Theta/2\pi$ times the lattice constant along the synthetic dimension.

\begin{figure}
\includegraphics[width=1\textwidth]{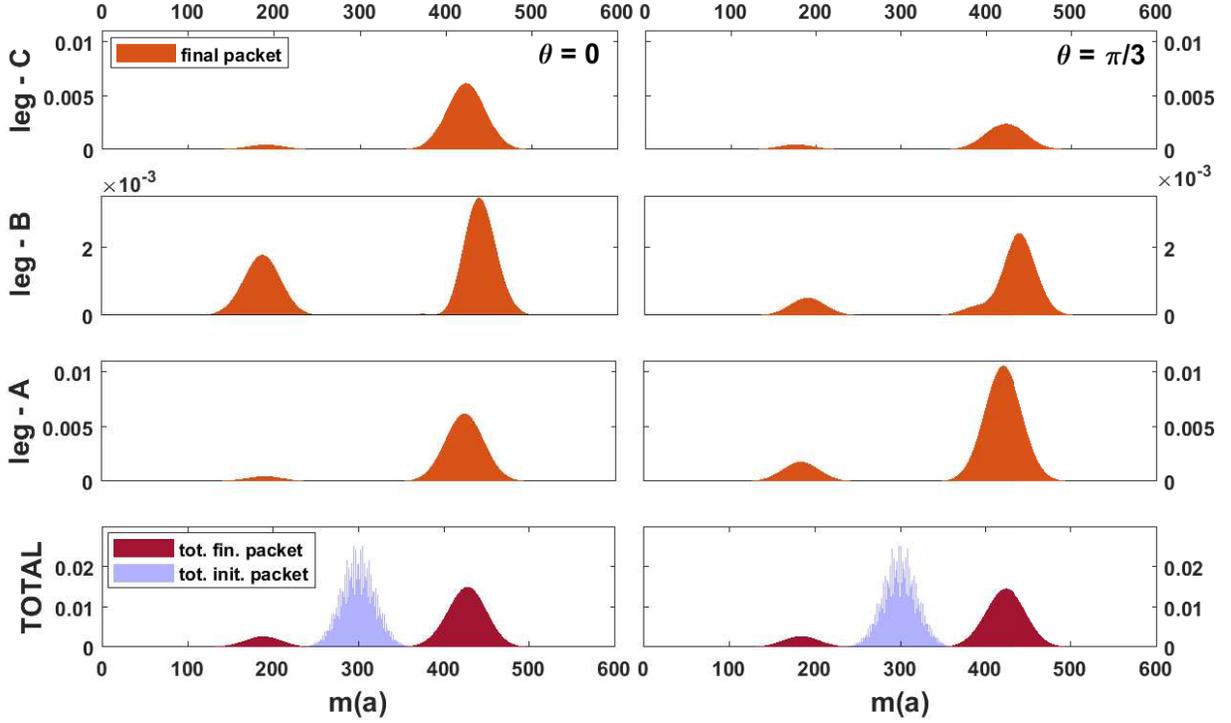}
\caption{The time evolution of the density at each leg for the quench $\phi = 1/2 \to 0$. The initial packet is prepared around an eigenstate of $H^{(2)}_{3L}$ under $\phi=1/2$ at $k_0 = 0.15 \pi$ and is quenched to $H^{\Theta}_{3L}$ for the gauge angles, $\Theta = 0$ and  $\Theta = \pi/3$. The difference in each leg is because of the absent electric field $\frac{\Theta}{a}\delta(t)$ needed for the dynamical gauge equivalence. Notice that the total density is same for both gauge angles.}
\label{fig:threeQuench0theta}
\end{figure}

We start with an initial state which is an eigenstate of $\hat{H}^{(2)}_{3L}$ with $\phi_1=1/2$ at $k= 0.15 \pi$ in the lowest band. We consider a quench to $\hat{H}^{\Theta}_{3L}$ with two different values of the gauge angle $\Theta=0$ and $\Theta=\pi/3$. In both cases the initial wavepacket splits into three, as expected for a three leg ladder with full translational symmetry. Furthermore, the total weights of the three packets are the same for both angles. The weights of the packets remain the same for not only two gauge angles but they are independent of the gauge angle. While this measurable property is independent of the gauge choice, another measurable property, namely the distribution of each split wavepacket along the synthetic dimension depends on it. In Fig.\ref{fig:threeQuench0theta} one can see that while the total weight summed over the three hyperfine states is the same in both cases, the hyperfine make-up of each packet is different.

The independence of the packet weight from the gauge angle $\Theta$ is clearly seen when one considers the missing scalar potential that would have acted during the quench to make all the $H^{\Theta}_{3L}$ equivalent. As the static gauge transformation linking different values of $\Theta$ is equivalent to a shift only along the synthetic dimension, the missing (delta function in time) electric field would have also acted only along the synthetic dimension. The quench to different values of gauge angle $\Theta$, correspond to different physical experiments in all of which the magnetic field is shut-off but different electric fields are applied along the synthetic dimension. Hence, the distribution in real space is not effected by the gauge angle choice. In Fig.\ref{fig:threeQuench0theta}, we give the hyperfine distribution of the lowest wavepacket as a function of the gauge angle, $\Theta$ under such a quench where the initial Gaussian packet prepared at quasimomentum $k = 0.15 \pi$ and under magnetic flux $\phi = 1/2$ is evolved under zero flux, $\phi = 0$. The total density of each leg is clearly dependent on the gauge angle or the missing scalar potential as a function of $\Theta$.

\section{Wide ladders and the robustness of the edge states}
It is not possible for synthetic dimension experiments to simulate systems with a fully extended synthetic dimension. However, if a large number of internal states are utilized, resulting simulation would be a good approximation to a higher dimensional system. The recent experiments carried out on atoms with high nuclear spins, such as with I = $7/2$, $9/2$\cite{SyntheticChiralEdge} allow the creation of wide ladders which closely resemble two dimensional systems. In this section, we investigate the effects of a flux quench in a fifteen leg ladder and contrast the behavior with continuum finite strip.
\begin{figure}
\includegraphics[width=.8\textwidth]{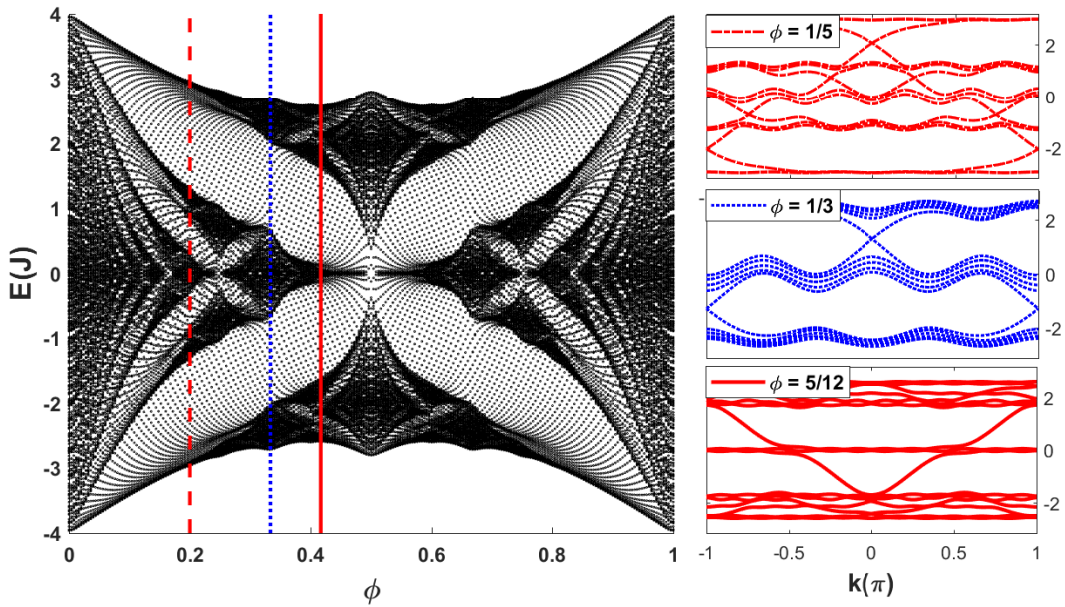}
\caption{(Color online) The energies of the fifteen-leg ladder sampled in 51 equally spaced points in k-space as a function of the magnetic flux, $\phi$. The energy spectrum is similar to the Hofstadter butterfly. We take three slices in this spectrum at $\phi = 1/5, 1/3, 5/12$ and investigate the quench scenarios for wavepackets made of edge states. The edge states reside within the energy gaps of the Hofstadter butterfly.}
\label{fig:fifteenLegEnergy}
\end{figure}

The behavior of a general wavepacket in a lattice with flux, formed around a momentum eigenstate with sufficient width in real space to allow for semi-classical motion, has two main features. The center of mass motion in real space (momentum space) will be modified by the Lorentz force due to the flux (Berry curvature), and the wavepacket will broaden as it is made up of superposition of states with neighboring momenta. The center of mass motion can be used as a direct probe of the underlying artificial magnetic field\cite{ChernMeastCoM}. A sudden quench has a moderate effect on the center of mass motion of a wavepacket, it evolves with the Lorentz force of the magnetic field after the quench. The effect of the quench on the packet shape is much more dramatic. After the quench the wavefunction ceases to be centred around a single momentum state in a single band but now is a superposition of many bands. As in the case of narrow ladders, the wavepacket splits into multiple packets, with different rates of dispersion.
\begin{figure}
\includegraphics[width=1\textwidth]{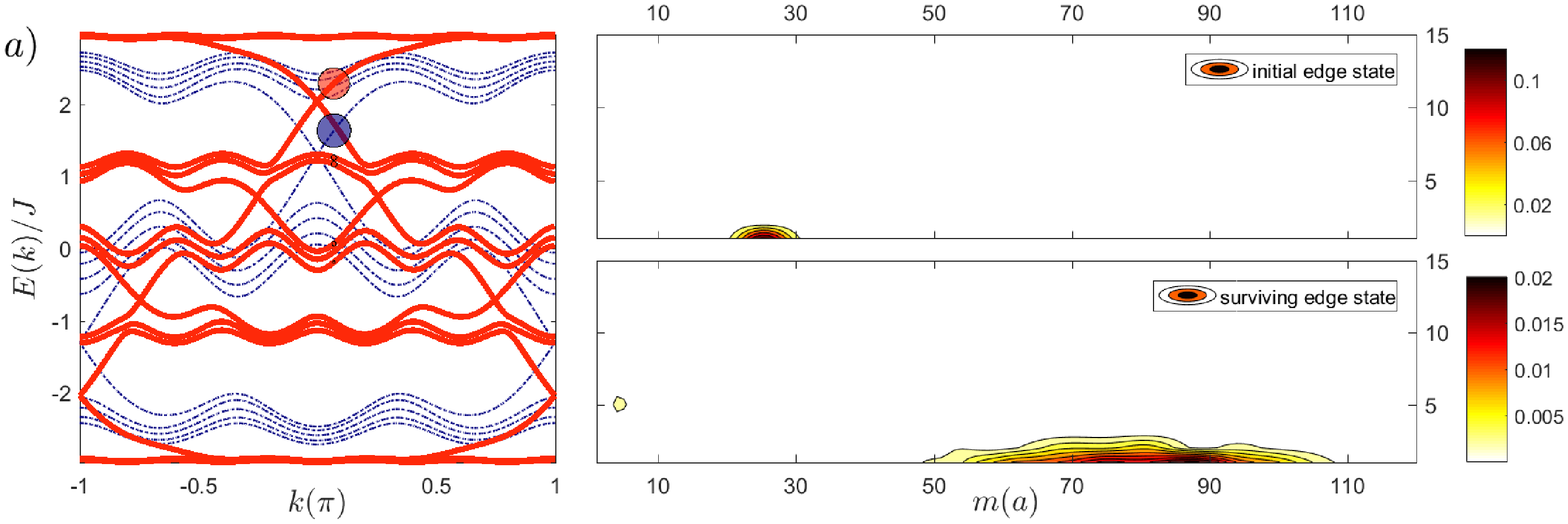}
\includegraphics[width=1\textwidth]{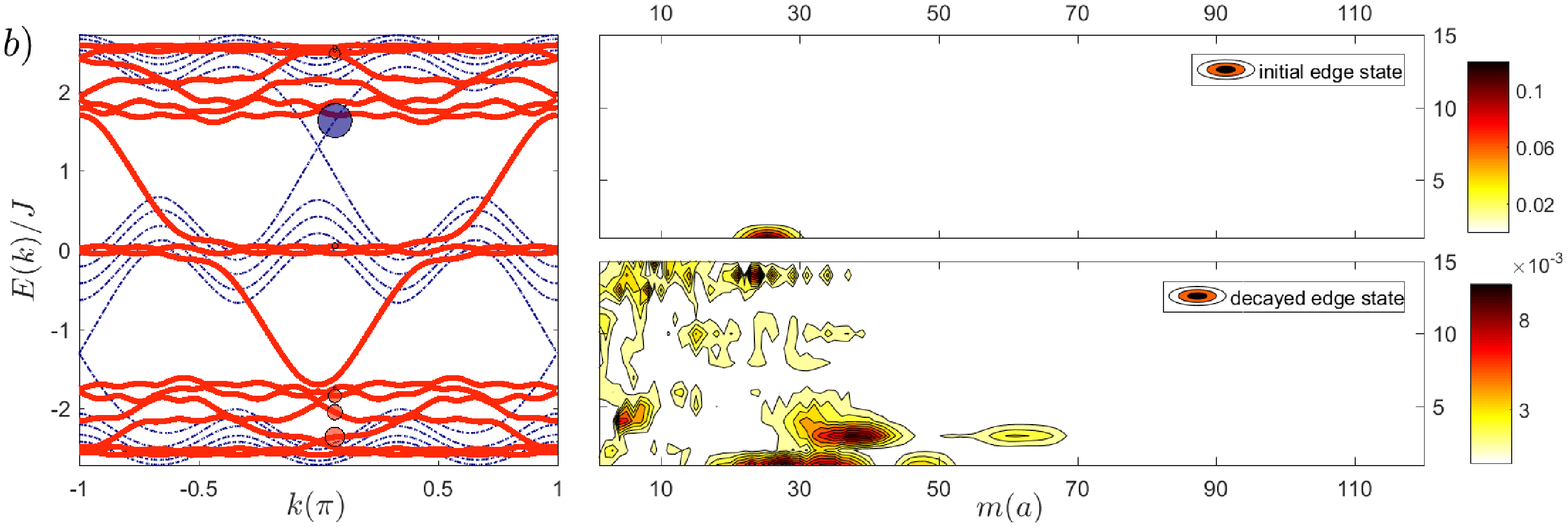}
\caption{Time evolved wavepacket of the edge state before and after the quench. The initial packet is prepared for $\phi = 1/3$ and the system is quenched into two magnetic flux values, $\phi = 1/5$ and $\phi = 5/12$, where the survival and the decay of the edge state is present. The upper (lower) plot on the left shows the weight of the initial packet (blue dot) on the final bands (red dots). The upper (lower) two stacked contour plots on the right are the time evolution of the wavepacket before and the after the quench. The edge state is prepared at $k_0 = 0.07 \pi$ and quenched as $\phi = 1/3 \to 1/5$. The same edge state is quenched as $\phi = 1/3 \to 5/12$. The sudden change in the flux redistributes the packet into the bulk.}
\label{fig:edgeStateRobustDestroyed}
\end{figure}

As an example, we consider the fifteen leg ladder. The energy spectrum for this ladder as a function of the flux per plaquette clearly shows how well this system resembles the two dimensional infinite lattice which has the Hofstadter butterfly spectrum (see Fig.\ref{fig:fifteenLegEnergy}). We first consider quench from $\phi_1=0$ to the final flux $\phi_2=1/3$, with an initial wavepacket centred in the middle with a width of 5 sites across the legs and 15 sites along the infinite direction with momentum along the infinite direction. After the quenches the splitting of the wavepacket is complicated by reflections from the edges and the wavepacket quickly broadens. This quick broadening upon the quench for bulk states was observed for different quench parameters and gauge choices, and we have not been able to discern any other universal behavior.
\begin{figure}
\includegraphics[width=1\textwidth]{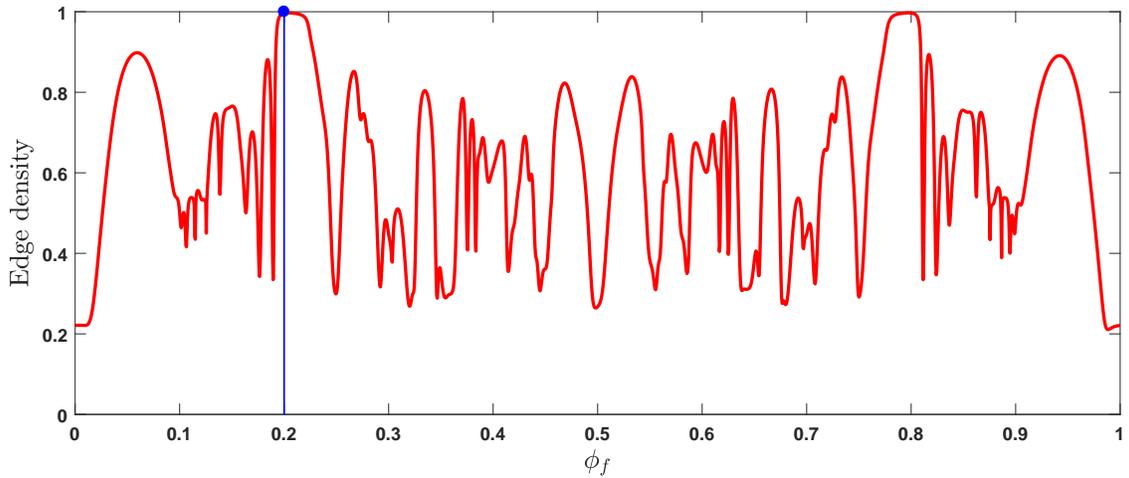}
\caption{(Color online) The edge density of an initial 15-leg ladder edge state prepared around $\phi=1/5$ and the lattice momentum $k = 0.01 \pi$ (blue line) after magnetic flux quench as $\phi: 1/5 \to \phi_f$, where $\phi_f \in [0,1]$. The edge density is calculated as the total density of the three legs at the edge.}
\label{fig:edgeDistribution}
\end{figure}

If the bulk system has a topological invariant the edge states become robust due to the bulk edge correspondence. It is natural to ask if any of this robustness can be retained in a quench. A general argument should not be expected for topological protection in a quench, however recent work has shown that, at least in two band systems, equilibrium topological invariants can be measured from dynamics after the quench \cite{HuiZhaiLinkingNum,DynaFleischhauer}. Here, we ask a simpler question and want to understand if the edge state remains bound to the edge or disperses into the bulk after the quench. Clearly both kinds of behavior are possible, a quench of an edge state from a finite flux to the zero magnetic field lattice where no edge modes exist results in decay into the bulk. Similarly, a small change in the flux carried out as a quench would not be expected to disturb the edge state.

We first take two quench scenarios for a right moving edge state packet prepared at $k= 0.07 \pi$ in magnetic flux $\phi_1= 1/3$. The magnetic flux is quenched as $\phi: 1/3 \to 1/5$ and $\phi: 1/3 \to 5/12$. The energy band diagrams for three flux values are shown in Fig.\ref{fig:fifteenLegEnergy}. The condition for the survival of the edge state under a flux quench is the existence of an edge state in one of the gaps at the same momentum of the final Hamiltonian. For example, the edge state at $k= 0.07 \pi$ in magnetic flux $\phi_1= 1/3$ transfers $88\%$ of its weight for the case $\phi: 1/3 \to 1/5$. The form of the edge states before and after the quench is shown in Fig.\ref{fig:edgeStateRobustDestroyed}. However, the edge state totally decays into the bulk states for the case $\phi: 1/3 \to 5/12$ as seen in Fig.\ref{fig:edgeStateRobustDestroyed}, because there is no edge state mode at this momentum for the quenched Hamiltonian (see Fig.\ref{fig:fifteenLegEnergy}).

We obtain a quantitative measure for the survival of the edge state following the quench by investigating the long time average of the density near the edge. After the quench the initial state (prepared at $\phi = 1/5$ and the lattice momentum $k = 0.01 \pi$) is expanded in terms of the new eigenmodes, however any crossterms in the density will oscillate with the frequency difference between the modes. As these oscillations (off-diagonal terms in the density matrix) average out to zero in the long time limit, we can calculate the residual density simply from the overlaps.  In Fig.\ref{fig:edgeDistribution} we give the post-quench edge density, defined as the total density in the first three sites in a 15 leg ladder as a function of the final flux. The crowded peak structure seen in the figure follows the existence of well defined gaps and corresponding edge states in this finite system. Thus, final fluxes close to rationals with small denominators are more prominent. In the zero flux and $1/2$ flux limits the spectrum does not have a well defined gap, hence the density at the edge falls to the uniform distribution limit of $3/15=0.2$.

\begin{figure}
\includegraphics[width=0.95\textwidth]{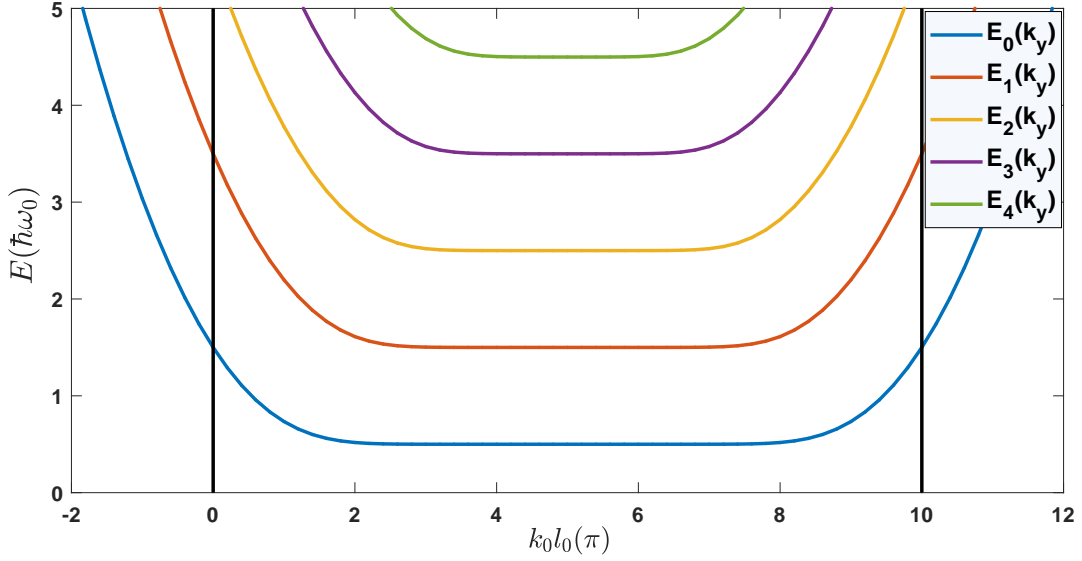}
\caption{The energy spectrum of a particle under a uniform magnetic field confined to $0<x<L = 10 \ell_0$ is numerically calculated. If the guiding center coordinate, $k_0$ is away from the edges, $0<k_0<10$, the energy has a flat spectrum and is made of degenerate Landau Levels. If the guiding center is outside the bulk, the energy spectrum depends highly on the guiding center, $k_y$ and the eigenfunctions become edge states.}
\label{fig:edgeenergy}
\end{figure}

As a natural extension of the wide ladders, we consider a continuum system and discuss the dynamics of a wavepacket in such a setting. We choose a strip of width $L$ in the $\hat{x}$ direction, infinite in the $\hat{y}$ direction
\begin{equation}
{\cal H}_{strip}= \frac{1}{2} \left( \vec{p} - \vec{A} \right)^2 + V(x,y),
\end{equation}
where
\begin{equation}
V(x)=\left\{ \begin{array}{r@{\quad:\quad}l} 0 & 0 < x  \leq L \\ \infty & x<0, x > L \end{array} \right.
\end{equation}
All parameters and functions are non-dimensionalized by the length scale $\ell_0 = \sqrt{\hbar/m \omega_c}$ and by the energy scale $E = \hbar \omega_c$, where $\omega_c = e B_0 /m$.
Choosing the vector potential in the Landau gauge preserves the translational invariance in the $y$ direction
\begin{equation}
\vec{A_1}= x \hat{y},
\end{equation}
and we only consider high magnetic fields where the magnetic oscillator length $\ell_0$ is much smaller than the strip width $L$. The canonical momentum in the $y$ direction $k_y$ is a conserved quantity which can be used to define the guiding center coordinate $x_g = k_y$. The energy spectrum as a function of the guiding coordinate is numerically calculated in Fig.\ref{fig:edgeenergy}, which shows two distinct types of eigenstates. If the guiding center coordinate is in the bulk of the strip, i.e. a distance of more than a few magnetic lengths away from the edges, the spectrum is flat forming the degenerate Landau levels. If the guiding center is outside the bulk region the corresponding eigenfunctions are edge states, with energies strongly depending the momentum in the $y$ direction.

We first demonstrate that the gauge choice is also relevant in the continuum by considering a wavepacket in the bulk.
Let us consider a wavepacket in the bulk, which is made up of eigenstates of this Hamiltonian all of which are in the lowest Landau level
\begin{eqnarray}
\psi(x,y)&=& \frac{1}{\sqrt{ \Delta \sqrt{\pi}}} \int_{-\infty}^{\infty} dk_y e^{-\frac{(k_y-k_0)^2}{2 \Delta^2}} \Psi_{0,k_y} (x,y), \\
\Psi_{0,k_y}(x,y) &=&  \frac{e^{i k_y y}}{\sqrt{2 \pi}} \frac{e^{-(x-k_y)^2/2}}{\pi^{1/4}},\quad E_{0,k_y} = 1/2.
\end{eqnarray}
where the center of the wavepacket is controlled by the average canonical momentum in the $y$ direction, $\langle \vec{x} \rangle = (k_0,0)$. The width in both directions are scaled with $\langle x^2 \rangle = 1 + \Delta^2$ and $\langle y^2 \rangle = (1+\Delta^2)/\Delta^2$. This packet is static, as it is formed by degenerate states all of which are in the lowest Landau level. If the Hamiltonian is suddenly changed so that the new artificial vector potential is
\begin{equation}
\vec{A_2}= (x-d) \hat{y},
\end{equation}
where $d$ is a constant, this wavepacket will no longer be static. Although both vector potentials define the same magnetic field, making the Hamiltonians before and after the quench equivalent up to a static gauge transformation, once again the missing scalar potential at the moment of the quench results in a momentum kick given to the wave packet. Time evolution of the wavepacket is easy to calculate as the initial state is essentially a coherent state of the new Hamiltonian,
\begin{eqnarray}
\psi(x,y,t)=   e^{-i \omega_c t /2} e^{-\frac{d^2}{4} (1- e^{-2i \omega_c t})} \frac{1}{\sqrt{\pi}} \sqrt{\frac{\Delta}{1+\Delta^2}}&& e^{i \Big[ (x - d (1-e^{-i \omega_c t}))y + \frac{y}{1 + \Delta^2} \big[ k_0 - x + d (1 - e^{-i\omega_c t}\big] \Big]} \\
&&e^{- \frac{\Delta^2}{1+\Delta^2}\frac{y^2}{2}} e^{- \frac{1}{2 (1+ \Delta^2)} \big[ x - ( k_0 + d (1 - e^{-i\omega_c t}))\big]^2}.
\end{eqnarray}
The absolute square of this function has a simpler form,
\begin{equation}
|\psi(x,y,t)|^2 = \frac{1}{\pi} \frac{\Delta}{1+\Delta^2} e^{-\frac{\Delta^2}{1+\Delta^2} \big[ y -d \sin{\omega_c t} \big]^2} e^{-\frac{1}{1+\Delta^2} \big[ x -(d+k_0) + d \cos{\omega_c t} \big]^2}.
\end{equation}
The wavepacket oscillates with the cyclotron frequency, and is no longer in the lowest Landau level. The excitation to higher Landau levels is a direct consequence of the energy imparted on the particle by the momentary electric field during the quench, $\vec{E} = \frac{d}{\tau} \hat{y}$ for the time dependent vector potential $\vec{A}(\vec{r},t) =  (x-d\frac{t}{\tau})\hat{y}$. The center of mass follows the classical trajectory,
\begin{equation}
\langle \vec{x}_{CoM}(t) \rangle = \Big( k_0 + d(1 - \cos{\omega_c t}), d \sin{\omega_c t} \Big)
\end{equation}

As expected the gauge dependence during the quench is not fundamentally different from the lattice scenarios considered in the previous sections.
\begin{figure}
\includegraphics[width=0.87\textwidth]{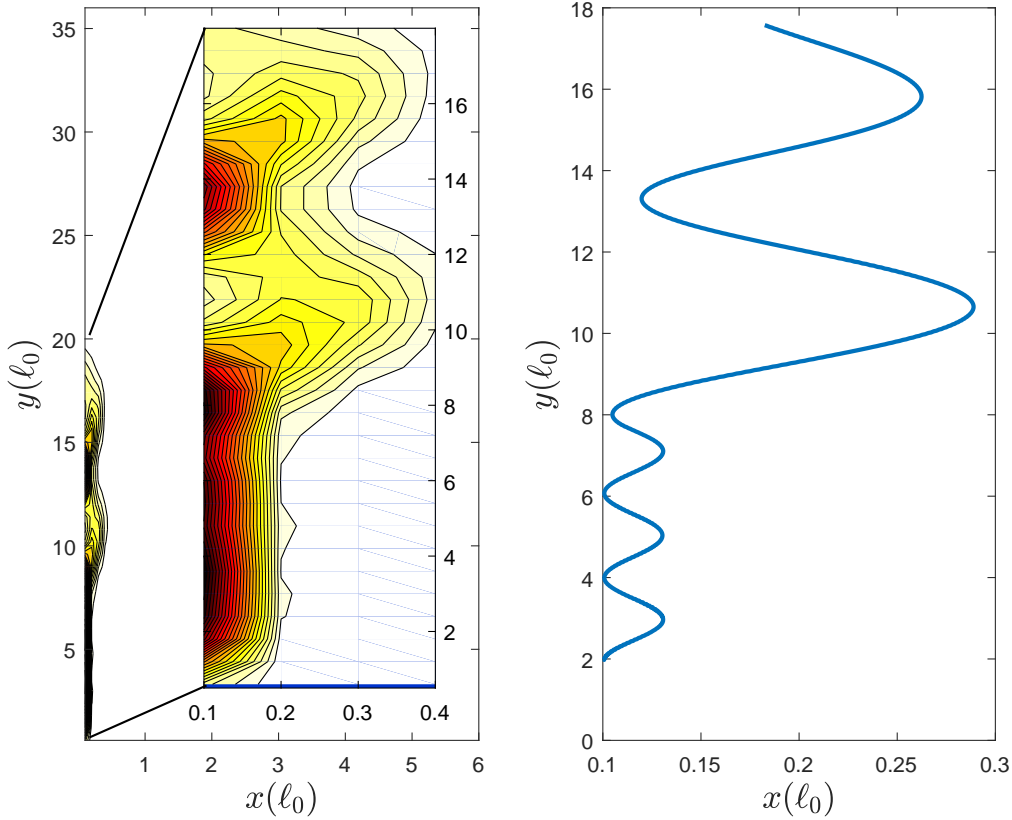}
\caption{The magnetic field quench response of an initial edge packet in the continuum limit. The initial packet is prepared as a slightly shifted version of the lowest eigenstate for the guiding center $k_0 = -2$ in the $x$ direction and Gaussian in $y$ direction. Hence, the initial packet have enough number of edge states to have a skipping orbit even before the quench.  The system is quenched as $B_1/B_0 = 90 \to 35 $, and $B_0$ is the magnetic field determines the length scale $\ell_0 = \sqrt{\hbar/eB_0}$, $B_1$ is the quenched magnetic field and energy scale $E_0 = \hbar e B_0 /m$ of the system. The center-of-mass position is plotted as a function of time. The pre- and post-quench packets clearly show skipping orbits. Depending on the magnitude of the final field, the skipping orbit has a different radius. Edge state is always robust as there are infinite number of available edge states in the continuum.}
\label{fig:ctnuum}
\end{figure}

For the edge states, the guiding center coordinate lies outside the strip. If the guiding center is more than a few magnetic oscillator lengths outside the strip, the effective one dimensional potential near the edge can be approximated by a linear potential. In this case, the wavefunction of an edge state is approximately
\begin{eqnarray}
\psi_{n,k_y} (x,y) &=& A_0 Ai \big( [2 k_y^2]^{1/3} (x-\frac{E_n}{k_y}+\frac{k_y}{2})\big) \frac{e^{i k_y y}}{\sqrt{2 \pi}},\\
E_{n,k_y} &=& \frac{k_y^2}{2} - \Big( \frac{k_y^2}{2} \Big)^{1/3} z_n,
\end{eqnarray}
$A_0$ is the normalization constant and $z_n$ are the zeros of the Airy function\cite{watson1995}.

Now consider a sudden change in the vector potential to
\begin{equation}
\vec{A_2}= \frac{B_1}{B_0} x \hat{y},
\end{equation}
which represents a change to a new magnetic field. The evolution of the state will be controlled by the overlaps of the initial wavefunction with the new eigenstates. If the significant overlaps are only with the bulk wavefunctions then the packet will disperse into the bulk, otherwise the wavepacket will remain bound to the edge. As the initial edge state is confined to within one magnetic length of the edge the overlaps can be estimated by the local density of states in the new Hamiltonian. Unlike the lattice problem there are infinitely many edge states in the continuum problem, and the local density of states near the edge are dominated by the new edge states. Thus, in almost all cases, excluding the total turn-off of the magnetic field, the edge state is robust and remains bound to the edge. In Fig.\ref{fig:ctnuum}, we plot the density of such a wavepacket after the quench and observe that an edge state before the quench becomes a skipping orbit, but remains bound to the edge\cite{Montambaux2011}. This behavior supports our finding in the lattice that the edge state is robust if there are protected edge states after the quench.

\section{CONCLUSION}
Throughout the paper we considered scenarios where the artificial magnetic field in a lattice system is abruptly changed. The most important conclusion that can be reached from our calculations is that the gauge choice in the implementation of the artificial magnetic field can have observable consequences in a quench. To be more specific, quenching into two Hamiltonians which can be related by a static gauge transformation gives physically different results for the same initial state.

We explored this gauge dependence first by considering a six site lattice model and calculating the dynamics during the magnetic field change. We have observed that if the magnetic flux change is carried out adiabatically, the dynamics is independent of the gauge choice. However for non adiabatic changes, including quenches, physical observables depend on specifically how the vector potential is implemented in the experiment. It is possible to recover full gauge symmetry only by implementing a time dependent scalar potential, which corresponds to time dependent on site energies. We believe both gauge choices we consider in Sec II can be implemented in a synthetic dimension experiment. One gauge choice corresponds to isolating two sites in real space by a superlattice potential and using spin-1 hyperfine levels as the synthetic dimension while the other gauge choice would need three real space lattice sites with two sites along the synthetic dimension.

We then investigated the effect of a sudden quench on a wavepacket in a two or three leg ladder. The wavepacket splits into smaller wavepackets moving with different velocities after the quench, but the number, the weights and and the velocities of these wavepackets depend on the gauge choice.  In particular, we considered quenches from a zero magnetic field ladder to two different implementations of the vector potential giving the same magnetic flux $\phi$. For vector potentials parallel to the n-leg ladder, the wave packet splits into at most $n$ smaller wavepackets. For the more experimentally relevant case, a vector potential parallel to the synthetic dimension the zero field packet splits into at most $n^2$ wavepackets, regardless of the value of the flux $\phi$. In both cases the  weights and the number of wavepackets are found from the overlaps of the initial state with the states at different bands at the same quasimomentum value.  We also considered a quenches from a finite magnetic flux to zero flux and investigated the effect of a trivial gauge choice corresponding to a constant vector potential parallel to the synthetic dimension. We find that such a constant vector potential does not affect the number of split packets, or their total weight, but it modifies the hyperfine state composition within the packet. We believe both kinds of experiments are within the reach of current experimental capabilities.

Finally, we considered the fate of topologically protected edge states upon a quench. In a wide ladder, the edge state might disperse quickly or remain mostly bound to the edge after the quench. The edge state remains bound only if most of its weight is transferred to an edge state in the new Hamiltonian. We find that the robustness of the edge state is controlled by the existence of an edge state at the same gap in the post-quench Hamiltonian. In the continuum the existence of edge states is guaranteed by the termination of Landau levels at the edge, and an edge state becomes a skipping orbit wavepacket after the quench.

We hope that our results stimulate further experimental interest in dynamics following artificial gauge field quenches.

\begin{acknowledgments}
F.Y. is supported by T\"{u}rkiye Bilimsel ve Teknolojik Ara\d{s}t{\i}rma Kurumu (T\"{U}B\.{I}TAK) Scholarship No. 2211.
M.\"{O}.O. acknowledges T\"{u}rkiye Bilimsel ve Teknolojik Ara\d{s}t{\i}rma Kurumu (T\"{U}B\.{I}TAK) project No 116T974.
\end{acknowledgments}
\appendix*
\section{N-LEG LADDER PACKET SPLITTING AFTER A FLUX QUENCH}\label{Appnd}
We consider a N leg ladder with the gauge choice, $\vec{A}_{||}(\vec{x}) = B_0 y \hat{x}$, the Hamiltonian is,
\begin{eqnarray}\label{HNparallel}
\hat{H}^{||}_{NL} &=& - J \sum_{m = -\infty}^{\infty} \Big[ \sum_{r = 0}^{n-1} \big( e^{-i 2 \pi \phi r} a^{\dagger}_{m+1,r} a_{m,r}  + H.c \big) + \sum_{r = 0}^{n-2} \big( a^{\dagger}_{m,r+1} a_{m,r} +  H.c. \big) \Big],
\end{eqnarray}
and the magnetic flux is $\phi = a^2 B_0$. Define the Fourier transform operators in the parallel direction,
\begin{eqnarray}
a_{k,r}^{\dagger} &=& \frac{1}{\sqrt{2 \pi}} \sum_{m=-\infty}^{\infty} a_{m,r}^{\dagger} e^{-i k m}, \quad -\pi \le k \le \pi, \nonumber \\
a^{\dagger}_{m,r} &=& \int^{\pi}_{-\pi} \frac{dk}{\sqrt{2 \pi}} a^{\dagger}_{k,r}  e^{i k m}.
\end{eqnarray}
The operators, $a_{m,r}^{\dagger}$ and $a_{k,r}^{\dagger}$ are the creation operators in the position and the momentum space. The Hamiltonian in Eq.\ref{HNparallel} in the momentum space is,
\begin{eqnarray}\label{HNparallelK}
\hat{H}^{||}_{nL} &=& \int^{\pi}_{-\pi} dk \sum_{r_1,r_2 = 0}^{N-1} a^{\dagger}_{k,r_1} A_{r_1,r_2}(k) a_{k,r_2},
\end{eqnarray}
and the matrix elements are
$$ A_{r_1,r_2}(k) = - J
\begin{pmatrix}
    2 \cos{(2 \pi \phi 0 - k)} & 1 & 0 & \dots  & 0 \\
    1 & 2 \cos{(2 \pi \phi 1 - k)} & 1 & \dots  & 0 \\
    \vdots & \vdots & \vdots & \ddots & \vdots \\
    0 & 0 & 0 & \dots  & 2 \cos{(2 \pi \phi (N-1) - k)}
\end{pmatrix}_{N \times N}$$.
In the diagonal basis,
\begin{eqnarray}\label{transF}
a_{k,r} = \sum_{s=0}^{N-1} U_{r,s}(k) b_{k,s},\\
b_{k,r} = \sum_{s=0}^{N-1} U_{s,r}^{*}(k) a_{k,s},\\
\end{eqnarray}
where $U^{\dagger} U = I$, the matrix $A_{r_1,r_2}(k)$ is diagonalized as $\sum_{r_1,r_2} U_{n_1,r_2}^{*}(k) A_{r_1,r_2}(k) U_{r_2,n}(k) = \varepsilon_{n_1}(k) \delta_{n_1,n}$. The Hamiltonian in the parallel gauge is,
\begin{eqnarray}\label{HNdiagK}
\hat{H}^{||}_{nL} &=& -J \int^{\pi}_{-\pi} dk \sum_{n=0}^{N-1} \varepsilon_{n}(k) b^{\dagger}_{k,n} b_{k,n}.
\end{eqnarray}

Our only requirement for the initial state is that it is an eigenstate of the translation operator in the absence of the magnetic flux,
\begin{equation}\label{initPsiNL}
|\psi_{k_0} (0) \rangle = \sum_{r=0}^{N-1} \psi_r(k_0) a^{\dagger}_{k_0,r}  | vac. \rangle.
\end{equation}
The coefficients $\psi_r(k_0)$ are chosen arbitrarily ensuring the normalization. Using Eq.\ref{transF}, the initial is expanded in the new basis with magnetic flux $\phi$,
\begin{equation}
|\psi_{k_0} (0) \rangle = \sum_{n=0}^{N-1} \left( \sum_{r=0}^{N-1} U^{*}_{n,r}(k_0) \psi_r(k_0) \right) b^{\dagger}_{k_0,n} | vac. \rangle = \sum_{n=0}^{N-1} W_n(k_0) b^{\dagger}_{k_0,n} | vac. \rangle.
\end{equation}
Note that $W_n(k_0)$ can also be written as $W_n(k_0)  = \langle k, n;\phi | \psi_{k_0}(0) \rangle$, where $|k, n;\phi \rangle = b^{\dagger}_{k_0,n} | vac. \rangle$. The time evolved form of this arbitrary wavefunction is,
\begin{eqnarray}\label{timEvoParall}
|\psi^{||}_{k_0} (t) \rangle &=& e^{-i \frac{t}{\hbar} \hat{H}^{||}_{NL}} \sum_{n=0}^{N-1} W_n(k_0) b^{\dagger}_{n,k_0} | vac. \rangle,\\
&=& \sum_{n=0}^{N-1} W_n(k_0) e^{i \frac{J t}{\hbar} \varepsilon_n(k_0)} b^{\dagger}_{n,k_0} | vac. \rangle.
\end{eqnarray}
Hence, an arbitrary wavefunction at lattice momentum $k_0$ {\bf splits into $N$ packets} when it is quenched to $ \hat{H}^{||}_{NL}$, where each packet  $n$ has weight $|W_n(k_0)|^2$.

We choose another set of gauge fields which leads the same observables in the static case, this perpendicular gauge is $\vec{A}_{\perp}(\vec{r}) = -B_0 x \hat{y}$. The relation between the two vector potentials are,
\begin{equation}
\vec{A}_{\perp}(\vec{x}) = \vec{A}_{||}(\vec{x}) - \vec{\nabla}\Lambda(\vec{x}), \quad \Lambda(\vec{x}) = B_0 x y = \phi m r,
\end{equation}
the gauge transformation, therefore, is
\begin{equation}
U = e^{i 2 \pi \phi \sum_{m,r} m r a^{\dagger}_{m,r} a_{m,r}}, \quad U^{\dagger} U = I.
\end{equation}
The corresponding form of $H_{NL}^{||}$ in the perpendicular static gauge is,
\begin{eqnarray}
H_{NL}^{\perp} &=& U H_{NL}^{||} U^{\dagger}, \nonumber \\
&=& - J \sum_{m = -\infty}^{\infty} \Big[ \sum_{r = 0}^{n-1} \big( a^{\dagger}_{m+1,r} a_{m,r}  + H.c \big) + \sum_{r = 0}^{n-2} \big( e^{i 2 \pi \phi m} a^{\dagger}_{m,r+1} a_{m,r} +  H.c. \big) \Big].
\end{eqnarray}

The same initial wavefunction in Eq.\ref{initPsiNL} is time evolved with $\hat{H}^{\perp}_{NL}$ as follows,
\begin{eqnarray} \label{timEvoPerp}
|\psi^{\perp}_{k_0} (t) \rangle &=& e^{-i \frac{t}{\hbar} \hat{H}^{\perp}_{NL}} |\psi_{k_0} (0) \rangle, \nonumber \\
&=& U e^{-i \frac{t}{\hbar} H_{NL}^{||}} U^{\dagger} |\psi_{k_0} (0) \rangle.
\end{eqnarray}
The first $U$ term in Eq.\ref{timEvoPerp} does not change the density, this modified phase on each lattice site is cancelled out. In addition, the number of wavepackets is $N$ times number of lattice momentum states in $U^{\dagger} |\psi_{k_0} (0) \rangle$. Note that interpretation is in parallel with Eq.\ref{timEvoParall} but this time $U^{\dagger} |\psi_{k_0} (0) \rangle$ is not restricted to single lattice momentum value. Investigating the term $U^{\dagger} |\psi_{k_0} (0) \rangle$,
\begin{eqnarray}
U^{\dagger} |\psi_{k_0} (0) \rangle &=& e^{- i 2 \pi \phi \sum_{m,r} m r a^{\dagger}_{m,r} a_{m,r}} \sum_{r^{'}=0}^{N-1} \psi_{r^{'}}(k_0) a^{\dagger}_{k_0,r^{'}}  | vac. \rangle , \nonumber \\
&=& e^{- i 2 \pi \phi \sum_{m,r} m r a^{\dagger}_{m,r} a_{m,r}} \sum_{r^{'}=0}^{N-1} \sum_{m^{'}=-\infty}^{\infty} \psi_{r^{'}}(k_0) e^{- i k_0 m^{'}} a^{\dagger}_{m^{'},r^{'}}  | vac. \rangle , \nonumber \\
&=& \sum_{r=0}^{N-1} \sum_{m^{'}=-\infty}^{\infty} \psi_{r}(k_0) e^{- i k_0 m^{'}}  e^{- i 2 \pi \phi m^{'} r} a^{\dagger}_{m^{'},r}  | vac. \rangle , \nonumber \\
&=& \sum_{r=0}^{N-1} \psi_{r}(k_0) a^{\dagger}_{k_0+2 \pi \phi r, r} | vac. \rangle.
\end{eqnarray}
We can expand each $a^{\dagger}_{k,r}$ operator in $b^{\dagger}_{k,r}$ basis using Eq.\ref{transF} as,
\begin{eqnarray}\label{UdagPsik}
U^{\dagger} |\psi_{k_0} (0) \rangle &=& \sum_{r=0}^{N-1} \psi_{r}(k_0) \left( \sum_{s=0}^{N-1} U_{s,r}^{*}(k_0+2 \pi \phi r) b^{\dagger}_{k_0+2 \pi \phi r,s} \right) | vac. \rangle , \nonumber \\
&=& \sum_{r=0}^{N-1} \sum_{s=0}^{N-1} c_{r,s}(k_0+2 \pi \phi r)b^{\dagger}_{k_0+2 \pi \phi r,s} | vac. \rangle,
\end{eqnarray}
and $c_{r,s}(k) = \psi_r(k) U^{*}_{s,r}(k)$.

Therefore, after the time evolution $e^{-i \frac{t}{\hbar} H_{NL}^{||}}$, there are {\bf at most $N^2$ packets} with the amplitude $|c_{r,s}(k)|^2$ and velocity $v_{s,r} = \frac{\partial \varepsilon_s(k)}{\partial k} \Bigg\rvert_{k_0+ 2 \pi \phi r}$ at $s$th band and the lattice momentum $k_0+2 \pi \phi r$.
There are no further restrictions other than the total number of bands, $N q$. If $N^2 > Nq$, the maximum number of packets are naturally restricted by $Nq$.

\bibliography{magnetic_flux_quench_v12}
\end{document}